\documentclass[manuscript]{aastex}
\usepackage[margin=1in]{geometry}
\usepackage{amsmath}
\usepackage{amssymb}
\usepackage{color}
\usepackage{starfont}
\usepackage{pdflscape}

\makeatletter
\let\@dates\relax
\makeatother

\slugcomment{Running Title: M DWARF FLARES ON HZ EARTH-LIKE PLANET}

\begin{document}

\title{Modeling Repeated M-dwarf Flaring at an Earth-like Planet in the Habitable Zone: I. Atmospheric Effects for an Unmagnetized Planet}

\author{\underline{Matt A. Tilley}$^{1,2,3,\ast}$, Ant\'{i}gona Segura$^{2,4}$, Victoria Meadows$^{2,3,5}$,  Suzanne Hawley$^{2,3,5}$, James Davenport$^{2,6}$}

\altaffiltext{$\ast$}{\underline{Corresponding Author:} University of Washington, Johnson Hall Rm-070, Box 351310, Seattle, WA 98195-1310, Ph.(cell): +1-206-484-5119, Email: mtilley@uw.edu}
\altaffiltext{1}{Dept. of Earth \& Space Sciences, University of Washington, Seattle, WA, USA}
\altaffiltext{2}{NASA Astrobiology Institute --- Virtual Planetary Laboratory Lead Team, USA}
\altaffiltext{3}{Astrobiology Program, University of Washington, Seattle, WA, USA}
\altaffiltext{4}{Instituto de Ciencias Nucleares, Universidad Nacional Aut\'{o}noma de M\'{e}xico, M\'{e}xico}
\altaffiltext{5}{Dept. of Astronomy, University of Washington, Seattle, WA, USA}
\altaffiltext{6}{Dept. of Physics \& Astronomy, Western Washington University, Bellingham, WA, USA}

\begin{abstract}
Understanding the impact of active M-dwarf stars on the atmospheric equilibrium and surface conditions of a habitable zone Earth-like planet is key to assessing M dwarf planet habitability. Previous modeling of the impact of electromagnetic (EM) radiation and protons from a single large flare on an Earth-like atmosphere indicated that significant and long-term reductions in ozone were possible, but the atmosphere recovered.  However, these stars more realistically exhibit frequent flaring with a distribution of different total energies and cadences. Here we use a coupled 1D photochemical and radiative-convective model to investigate the effects of repeated flaring on the photochemistry and surface UV of an Earth-like planet unprotected by an intrinsic magnetic field.  As input, we use time-resolved flare spectra obtained for the dM3 star AD Leo, combined with flare occurrence frequencies and total energies (typically 10$^{30.5}$ to 10$^{34}$ erg) from the 4-year Kepler light curve for the dM4 flare star GJ1243, with varied proton event impact frequency.   Our model results show that repeated EM-only flares have little effect on the ozone column depth, but that multiple proton events can rapidly destroy the ozone column. Combining the realistic flare and proton event frequencies with nominal CME/SEP geometries, we find the ozone column for an Earth-like planet can be depleted by 94\% in 10 years, with a downward trend that makes recovery unlikely and suggests further destruction. For more extreme stellar inputs O$_3$ depletion allows a constant  $\sim$0.1-1 W m$^{-2}$ of UV-C at the planet’s surface, which is likely detrimental to organic complexity. Our results suggest that active M dwarf hosts may comprehensively destroy ozone shields and subject the surface of magnetically-unprotected Earth-like planets to long-term radiation that can damage complex organic structures. However, this does not preclude habitability, as a safe haven for life could still exist below an ocean surface. 
 
\end{abstract}


\section{Introduction}

Whether M dwarf planets are good targets in the search for life is still an open problem of high complexity. There exist several factors that make such planets promising targets - primarily detectability and abundance. Rocky planets are easier to detect around these low mass stars (0.5 to 0.08 M$_\odot$) with current techniques and instruments. Red dwarfs are abundant in the solar neighborhood, they spend $\sim 10^{10}$ years in the main sequence, and biosignatures like N$_{2}$O and CH$_4$ may be more abundant and potentially easier to detect \citep[see reviews by][]{tarteretal2007,scaloetal2007,shieldsetal2016}. The abundance of M dwarf hosts, and therefore potentially habitable planets, is beneficial since planets orbiting these hosts in their proximal habitable zones encounter conditions that could make habitability challenging. 

Three dimensional planetary atmospheric models have dispelled concerns of atmospheric collapse by extreme temperature differences between the illuminated and dark hemispheres \citep[e.g.,][]{Dole1964} for tidally-locked planets orbiting M dwarf hosts in the habitable zone; the models applied to this problem showed that atmospheric circulation can distribute the stellar energy, lowering the temperature difference \citep[e.g.][]{joshi1997,joshi2003}. Active M dwarf host stars, however, exhibit magnetic activity originated by the interaction of their atmospheres with their magnetic fields driven by their mostly or totally convective interiors. One consequence of magnetic activity are flares, unpredictable releases of energy that range from "microflares" ($\sim 10^{29}$ ergs as measured in the U band, $\sim$3320-3980 \AA) to high energy flares with total energies as large as 10$^{34}$ ergs \citep[e.g.,][]{hawley1991,hawley2014}. Since the habitable zone around these stellar hosts is located at a fraction of an astronomical unit (AU), the energy flux from these events that impact potentially habitable worlds is increased by at least an order of magnitude compared to Earth.

Active M dwarfs exhibit flare activity with widely varying flare energies and frequencies, but generally following a power-law probability distribution \citep[e.g.,][]{hilton2011,hawley2014}. The most active of these dM3-dM5 stars have been observed to produce dozens of flares per day with total energies $\geq$10$^{30}$ erg, and potentially catastrophic events like the Great Flare of ADLeo at 10$^{34}$ erg \citep{hawley1991} at a frequency of once per month. Compare this to average solar activity with the highest frequency of a single event per day of energy $\sim$10$^{27}$ erg, up to approximately one 10$^{31}$ erg event per year \citep[e.g.,][]{crosby1993}. One of the largest observed solar flare events, known as the Carrington Event of 1859, had an estimated total energy of $\sim$10$^{32}$ erg \citep[e.g.,][]{cliver2013}, a full two orders of magnitude lower than the most energetic events observed on active M dwarfs. 

There are two aspects of flare events that could negatively impact the habitability of a planet by altering the atmospheric composition: increase in electromagnetic flux, and an eruption of energized charged particles. The energy injected by a flare into the stellar atmosphere results in a rise and peak of photometric brightness, known as impulsive phase. A gradual flare decay phase follows where the energy input decreases until the star slowly returns to quiescent state. During flare events the luminosity of the star in the X-ray, UV and visible increases up to 3 orders of magnitude \citep{scaloetal2007}, which can photochemically alter the upper atmosphere of a planet. For a planet that does not have atmospheric constituents (e.g., O$_3$, CO$_2$) to absorb this short wavelength radiation, the surface could be regularly irradiated. 

The second impactful aspect of flare events involves stellar energetic particles (SEP) that can be accelerated during the impulsive phase of flares. Our knowledge of these events comes only from solar correlations, as we have no method by which they can be remotely observed. Their effect on the atmosphere of a habitable planet likely depends on several factors: the particle energies, presence and orientation of a planetary magnetic field, and the chemical composition of the atmosphere. The probability of a proton event being associated with a stellar flare is dependent on the flare energy, with solar microflares ($\leq$10$^{27}$ erg) only rarely producing weak particle events, but the chance for a large flux of energetic particles reaches a near certainty for flares $\sim$10$^{28.3}$ erg (GOES class X2) and above \citep[e.g.,][]{yashiro2006,hudson2011,dierckxsens2015}. While not every proton event will hit a potentially habitable planet due to the combined geometry of the active regions of the star and the planetary orbit \citep[e.g.,][]{khodachenko2007}, M dwarf planets we consider are under significant threat. These planets orbit at fraction of an AU, and the stellar hosts exhibit high frequency, high energy flare events indicating the effects of multiple impacting proton events could be significant.

Studies on the effects of space weather on habitability have become crucial with the discovery of planets such as Proxima Centauri b and the TRAPPIST-1 systems' multiple habitable zone planets. Planetary atmospheric escape can be caused by XUV radiation that heats the planet´s exosphere driving a hydrodynamic wind that carries away the planetary atmosphere \citep[e.g.][]{lugerbarnes2015}; non-thermal processes such as the increased polar wind and ion pickup escape can potentially remove vast quantities of heavy ions such as N$^+$ and O$^+$ from the atmosphere \citep[e.g.,][]{ribas2016,airapetian2017,garcia2017};  increased forcing of the magnetosphere by either steady state or transient stellar wind events can also be impactful for atmospheric loss, and the subsequent atmospheric outflow \citep[e.g.,][]{dong2017,garraffo2016,garraffo2017}. However, there is so far less effort into studying the effects of space weather events on the chemical evolution of planets that retain their atmosphere, and how habitability might be affected in such instances.

\citet{segura2010} performed the first study for understanding the effect of a single, high energy flare (and associated SEP) on the atmospheric chemistry of a habitable planet; their study investigated the impact of an event equivalent to the 1985 April 12 flare from the M dwarf AD Leonis (AD Leo) \citep{hawley1991} on an Earth-like planet located within its habitable zone. The planetary atmosphere had a similar composition to that of present Earth (0.21 O$_2$, 1 bar surface pressure) and was located to receive the same integrated stellar flux as our planet. They simulated the effects from both UV radiation and protons on the atmospheric chemistry of a hypothetical habitable planet with a 1-D photochemical model coupled to a 1-D radiative/convective model. Observations available for this flare included UV spectroscopy (1150 - 3100 \AA) and optical spectroscopy (3560 - 4440 \AA), but no evidence of SEPs were available. To include particles, \citet{segura2010} used a relationship found for solar X-ray flare intensity and proton fluxes \citep{belov2005}. X-rays flare intensity were obtained from the Neupert effect, an empirical relation between the flare energy emitted in the UV and the X-ray peak luminosity \citep{mitra-kraevetal2005}. They estimated a proton flux associated with the flare of $5.9 \times 10^8$ protons cm$^{-2}$ sr$^{-1}$ s$^{-1}$ for particles with energies $>10$ MeV. Then they calculated the abundance of nitrogen oxides produced by the flare by scaling the production of these compounds during a large solar proton  event called the Carrington event \citep[e.g.]{cliver2013}. 

\citet{segura2010}'s results indicated that the UV radiation emitted during the flare does not produce a significant change in the ozone column depth of the planet. When SEPs were included, the ozone depletion reached a maximum of 94\% two years after the flare for a planet with no magnetic field. At the peak of the flare, the calculated UV fluxes that reached the surface, in the wavelength ranges that are damaging for life, exceed those received on Earth during less than 100 s. The atmospheric column then recovered and re-equilibrated within $\sim$50 years. \citet{segura2010} concluded that flares may not present a direct hazard for life on the surface of an orbiting habitable planet. 

Recent studies have been performed to determine the effect of flare electromagnetic energy on the habitability of planetary systems. The TRAPPIST-1 system has been investigated by \citet{vida2017} in a recent {\it Kepler/K2} study showing at least $\sim$0.75 cumulative flares per day of energy between 1.26$\times$10$^{30}$-1.24$\times$10$^{33}$ erg. This could produce detrimental UV flux at the potentially habitable planetary surfaces of TRAPPIST-1 b-h. \citet{omalley2017} investigated the potential UV related surface habitability of the TRAPPIST-1 system, and found that the oxic state of the atmosphere is key to protecting the surface, with even a thin-oxygen atmosphere ($\sim$0.1 bar) sufficient to keep UV-C from reaching the surface with any intensity. \citet{estrela2017} have identified superflares in the Kepler-96 system up to $\sim$1.8$\times$10$^{35}$ erg, and simulated the effects on both Archean and present-day Earth-like atmospheres, and found that the presence of an O$_3$ layer is crucial to protection of life under such highly irradiating events. 

Aside from \citet{segura2010}, these previous studies do not investigate the effects of proton events associated with stellar magnetic activity, nor the resilience and evolution of the O$_3$ column on a potentially habitable planet to such events. AD Leo, the star used for the \citet{segura2010} study, is one of the most magnetically active M dwarfs known. In the past few decades, only flares from the most active M dwarfs were studied because the UV emission from mid- and low-active red dwarfs fall below the detection threshold of the available instruments. Observations performed using {\it Hubble Space Telescope} showed that UV emission from chromospheric activity was also present on those stars usually classified as non-active \citep{walkowiczetal2008,franceetal2012,franceetal2013}, while the {\it Kepler/K2} mission \citep[e.g.,][]{hawley2014,davenport2014} and the {\it MOST} instruments \citep{davenport2016} showed a more detailed view on flares frequency from low mass stars. 

The present work extends the results from \citet{segura2010} to determine the effect of multiple M dwarf flares and energetic proton events on a potentially habitable world by taking advantage of the more recent observational campaigns noted above. We have updated the models used in \citet{segura2010} (see \S~\ref{sec:segura_ext}) to study the effect of multiple events on an atmosphere with similar composition of present Earth to determine the potential effects on the O$_3$ column and related biologically relevant UV flux at the surface of the planet. 

\section{Methods}

\subsection{Improvements extending \citet{segura2010}}
\label{sec:segura_ext}

A modified version of the 1D coupled radiative-convective and photochemical model used by \citet{segura2010} is used in in the present work. The radiative-convective climate model itself is a hybrid of two models: 1) a $\delta$ two-stream scattering algorithm that is used to calculate fluxes and uses correlated-$k$ coefficients to parameterize absorption by important atmospheric species, e.g.,  O$_3$, CO$_2$, H$_2$O, and CH$_4$; 2) for thermal-IR wavelengths, the rapid radiative transfer model (RRTM) implemented by \citet{segura2003} was used. The RRTM uses 16-term sums in each spectral band where $k$-coefficients are calculated to give high spectral resolution where Doppler broadening is important. 

The photochemical model solves 217 reactions that link 55 chemical species. Photolysis was calculated using a $\delta$ two-stream routine that allowed scattering between molecular gases and the included aerosol species \citep{segura2003}. Timesteps were solved using an implicit reverse Euler method, with initial timestep set to 10$^{-4}$ s, with increasing magnitude as the system reaches equilibrium. 

The coupling layer allowed the climate and photochemical model to cross-communicate and synchronize the atmospheric temperature structure, H$_2$O profiles and chemical alterations made for each timestep. The pressure layers calculated in the radiative-convective model were interpolated to the fixed altitude structure in the photochemical model, and then back, during the coupling procedure. All other details and inputs to the model not discussed below are identical to those discussed in \S 2-3 in \citet{segura2010}.

In the present work, we made four important modifications: \S~\ref{sec:segura_unfixed} describes how we unfixed the mixing ratios for three major atmospheric constituents (N$_2$, O$_2$, and CO$_2$) to achieve a more realistic mass balance; \S~\ref{sec:kepler_multi} details the treatment used for multiple flare events and flare light curve evolution based on recent observations; \S~\ref{sec:proton_events} discusses the considerations made regarding the interaction between multiple stellar proton events and the upper atmosphere; \S~\ref{sec:conundruum} discusses the extension of the stellar optical-NIR spectrum during flare events.

\subsection{Unfixed mixing ratios for O$_2$, CO$_2$, N$_2$ and Henry coefficients}
\label{sec:segura_unfixed}

For the single flare work addressed in \citet{segura2010}, the volume mixing ratios, {\it f}, of major atmospheric constituents O$_2$, CO$_2$, and N$_2$ were fixed at values corresponding to values of 0.21, 3.55$\times$10$^{-4}$, and 0.78, respectively. In the present work, these mixing ratios are enforced only at the surface layer of the planet, and all layers of the atmosphere above the surface were unfixed --– allowing these species to photochemically respond to the stellar inputs. 

Of particular note are the responses of O, O$_2$ and N$_2$ during the creation of NO$_x$ species generated by proton events. Previous studies have found that each ion pair (positive/negative species) created in the upper atmosphere from precipitating proton ionization events results in the production of 1.27 N atoms, particularly N(4S) and N(2D) \citep{porter1976}, and a NO$_x$ production rate of 1.3 to 1.6 per pair \citep{rusch1981}. The generation of NO$_x$ is driven by the dissociation of N$_2$ into the constituent, excited N-atoms which then react with O$_2$ to produce NO and O: 

\begin{align}
    \mathrm{N}(4S) + \mathrm{O}_2 &\rightarrow \mathrm{NO + O}	    \tag{R1} \label{eq:no_production1}\\
    \mathrm{N}(2D) + \mathrm{O}_2 &\rightarrow \mathrm{NO + O}		\tag{R2} \label{eq:no_production2}
\end{align}

\noindent and O$_3$ reacts with the produced NO to generate NO$_2$:

\begin{align}
    \mathrm{NO + O}_3 &\rightarrow \mathrm{NO}_2 + \mathrm{O}_2.  \tag{R3}\label{eq:no2_production}
\end{align}

Following the method of proton injection in \citet{segura2010}, we directly modify the atmospheric column to simultaneously include the NO and NO$_2$ profiles at or near the peak of the flare events; the excited N atoms are not directly injected. The amount of NO$_x$ directly input into the upper atmosphere to emulate production by proton precipitation (and therefore pair production) sets the amounts of N$_2$ and O$_2$ removed by the formulating reactions (as well as O$_3$, and the addition of O). For each NO molecule we inject, 0.5 N$_2$ and 1 O$_2$ molecules were immediately removed from the atmospheric column, and one O atom was added. For the case of NO$_2$ molecules injected at a specific altitude, the abundances are predicate on the previous formation of 
NO, as seen in Eq.~\ref{eq:no2_production}; during the formation of NO$_2$ from NO, 1 O$_3$ molecule is removed, and per the reactant NO that drove the production of the NO$_2$, 0.5 N$_2$ molecules are removed from the atmospheric column, and 1 O atom is added. Note that the product O$_2$ from Eq.~\ref{eq:no2_production} is cancelled by the assumed pre-production of the NO molecule, which consumes an O$_2$ as in Eqs.~\ref{eq:no_production1}~and~\ref{eq:no_production2}. For simulations containing multiple proton events, we limit the maximum NO (NO$_2$) injected into the atmosphere by the amount of available reactant O$_2$ (O$_3$) in the atmospheric column.

As a consequence of unfixing the mixing ratio of CO$_2$, its abundance was increased in the stratosphere by methane oxidation \citep{yung1999photochemistry}, as we will show in Sec.~\ref{sec:discuss_obs}.

Rainout and saturation for the new unfixed species were also modified to be temperature dependent, and the appropriate Henry coefficients are shown in Table~\ref{tab:henry_coeff} along with the initial mixing ratios at the surface level of the atmospheric column. All of the above changes to the 1D model did not significantly alter O$_2$ and N$_2$ during steady-state from their uniform altitude mixing ratio profiles of 0.21 and 0.78, respectively.

\subsection{Multiple flare events from Kepler observations}
\label{sec:kepler_multi}

\begin{figure}
\centering
\includegraphics[width=0.75\textwidth]{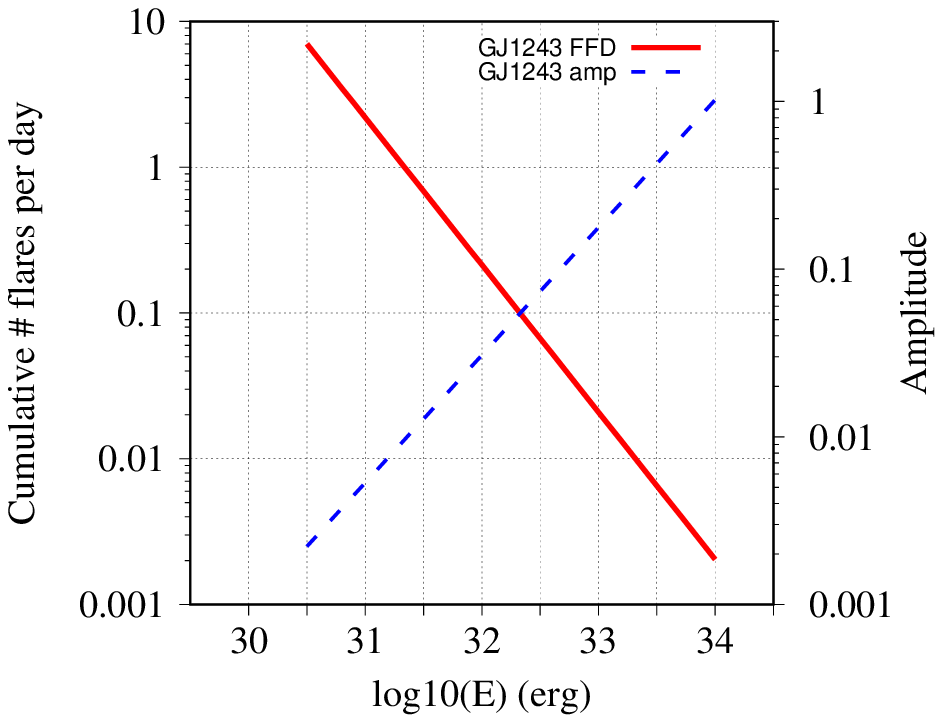}
\caption{The flare frequency distribution (FFD) and amplitudes observed from GJ1243 - used to generated flare distributions in the present work.}
\label{fig:ffdamp}
\end{figure}

To address the effects of multiple flares on the atmosphere, we built upon the ADLeo flare template used in the previous work of \citet{segura2010} by using statistical results from Kepler observations of the M4 star GJ1243. We incorporated the observations of the flare energies, durations, and amplitudes from GJ1243 \citep{hawley2014}, which included analysis of over 6100 GJ1243 flares measured over 11 months at 1-minute cadence. These observations also resulted in the empirical flare light curve evolution template we employ \citep{davenport2014}. The following are scaling relations derived from \citet{hawley2014} that we used to determine flare amplitude (relative increase in U-band flux), flare frequency, and event durations:

\begin{align}
	\log_{10}\frac{\Delta F_U}{F_U} = 0.7607\,\log_{10} E - 25.855 \label{eq:amplitude}\\
	\log_{10} \nu = -1.01\,\log_{10} E + 31.65 \label{eq:frequency}\\
	\log_{10} t = 0.395\,\log_{10} E -9.269 \label{eq:duration}
\end{align}

\noindent where $F_U$ is the stellar flux in the Johnson U-band, E represents the flare energy in ergs, $\nu$ is the flare frequency given in cumulative flares per day, and $t$ is the duration of the flare events in seconds which ranges from $\sim$600 s to $\sim$14,500 s for flares of energy 10$^{30.5}$ erg and 10$^{34}$ erg, respectively. Fig.~\ref{fig:ffdamp} shows a plot of the flare frequency and amplitudes as a function of the energy used to generate our flare distributions.

The present work explores several separately generated distributions of stellar flare events with duration of one month, six months, one year, 10 years, and 15 years. We simulated flares at a rate of $\sim$7 per day with energies in the Johnson U-band from a minimum of 10$^{30.5}$ erg up to a maximum of 10$^{34}$ erg, which occurs with frequency of 1 flare per $\sim$489 days according to the flare frequency distribution (FFD) for GJ1243 reported in \citet{hawley2014}, and shown in Fig.~\ref{fig:ffdamp}. An example extended, temporal flare distribution with $\sim$1277 (2555) flares over six months (one year) was generated from the GJ1243 FFD, and can be seen in Fig.~\ref{fig:dav_ffdflares}. As seen in Fig.~\ref{fig:ffdamp}, a relative flare amplitude of $\sim$1.04 corresponds to a flare with energy 10$^{34}$ erg - approximately the energy of the ‘great flare’ of ADLeo reported by \citet{hawley1991} - and $\sim$2.25$\times$10$^{-3}$ corresponds to a flare of size 10$^{30.5}$ erg from GJ1243. 

\citet{cliver2013} estimate the solar Carrington event of 1859 to be on the order of 5$\times$10$^{32}$ erg, events which by Eq.~\ref{eq:amplitude} have a relative amplitude of $\sim$0.075. However, any reference to 'Carrington-size' flares in the present work were calibrated by proton fluence ($\sim$1.1$\times$10$^{10}$ cm$^{-2}$), rather than total flare energy, and correspond to a total energy of $\sim$10$^{31.9}$ erg, with a flux amplitude of $\sim$0.0223. Flare flux was allowed to stack, enabling the simulation of complex flaring events as seen in the inset of the top panel in Fig.~\ref{fig:dav_ffdflares}.

The flare light curve template was also modified in the present work. Following \citet{davenport2014}, we constructed each flare event using a light curve template granting a rapid impulsive phase, and extended decay phase. Fig.~\ref{fig:lightcurves_comp} shows a relative comparison between the light curve shape used in \citet{segura2010} and an equivalent 10$^{34}$ erg flare from the present work for the first 6000 seconds of the flare events. It is worth noting that the difference between peak times results mostly from how one defines where a flare starts; with the template applied, we start from a much smaller increase in magnitude for the first timestep compared to \citet{segura2010}.

Integrating the impulsive phase from the start of the event to the peak for each flare results in a factor of $\sim$0.71 in flux delivered for the treatment in the present work treatment (solid blue) from \citet{davenport2014}, compared to results from the \citet{segura2010} light curve (dashed red). Integrating each flare from start to end, however, results in the present treatment depositing approximately twice the overall flux when compared to the treatment in \citet{segura2010}. Integrating both curves across the FWHM results in a factor of $\sim$1.72 increase of the flux compared to the original treatment in \citet{segura2010}. In the present work, the majority of the flare energy is delivered in the decay of the flare evolution due to the long timescale compared to the impulsive phase. It is worth noting that for flare energies $<$10$^{34}$ erg, the FWHM, rise and decay times (and therefore total duration, as seen in Eq.~\ref{eq:duration}) all occur at reduced scale with the decreased energies.

The per-flare timestep resolution was increased in our implementation to be more granular, from 20 flare points used in the 2010 work, to 30 points (12 in the rise phase instead of 9, 18 in the decay phase instead of 11) to obtain proper coverage of temporal flare morphology. The resulting flare in Fig.~\ref{fig:lightcurves_comp} is ~5.6$\times$ the duration of the flare used in \citet{segura2010} ($\sim$14,500 s versus $\sim$2,586 s) due to the long decay phase from the empirical template adopted from \citet{davenport2014}. This duration varies with the flare energy according to Eq.~\ref{eq:duration}. 

\subsection{Proton Event Scaling and Impact Probability}
\label{sec:proton_events}

In addition to UV radiation, flare events can produce proton events that can drive changes in atmospheric chemistry. In this section, we explain our process and reasoning behind the scaling of proton events for the flare energy ranges simulated, and the likelihood of their impact on an orbiting planet. Essentially, we follow the scaling method of \citet{segura2010} relying on a correlation between flare x-ray flux and proton fluence. This method matches well with observations at Earth during a particularly well-studied energetic solar event in 1989'; we expand the method to the orbital and stellar parameters for our extrasolar system of interest.

Scaling x-ray intensity is used to estimate proton fluence from flare events. We assume the spectral x-ray energy density in the 1-8 \AA\ bandpass reported for the 10$^{34}$ erg flare in \citet{hawley1991} scales directly with our relative flare flux amplitude relation in Eq.~\ref{eq:amplitude}. The relationship between proton fluence and relative flux or flare energy is shown in Fig.~\ref{fig:fluence_vs_flux}. \citet{belov2005} showed a correlation between X-ray intensity and proton flux for solar events observed over three solar cycles, using the IMP-8 and GOES satellite instruments. The GOES satellites orbit at geosynchronous altitudes, implying the GOES measurements of proton flux are typically taken from within the Earth's magnetosphere, except during strong flow due to a fast solar wind stream or the impact of a coronal mass ejection (CME). By this reasoning, the scaling relation we use here from \citet{segura2010} and \citet{belov2005} is based upon the magnetized Earth applied to our unmagnetized extrasolar planet, so it can therefore be assumed to be a lower bound of proton fluence for the latter.

To ensure our scaling is somewhat reasonable, we apply it and compare to an observed solar event at Earth. Keeping in mind the variability for solar and magnetospheric dynamics, we compare here our proton event fluence scaling with the well-studied solar proton event of the Oct 19/20 1989 flare that is classified as a GOES X13 event, i.e., peak X-ray flux 1.3$\times$10$^{-3}$ W m$^{-2}$ in the 1-8 \AA~bandpass as measured by the GOES-6/7 instruments \citep[e.g.][]{jackman1993,jackman1995,verronen2002}. The GOES satellites measured a peak of $\sim$40,000 proton flux units (pfu, particles sr$^{-1}$ cm$^{-2}$ s$^{-1}$) for protons with energy $\geq$10 MeV during an event which had duration of $\sim$27 hours. At the time of peak pfu measurement, the GOES satellites were well inside the Earth's magnetosphere.

We follow the scaling used in \citet{segura2010} based on \citet{belov2005}, and find that a GOES X13 flare would produce ~24,611 pfu at Earth which seems reasonable (if a bit underscaled) when compared to the peak GOES-6/7 measurements of $\sim$40,000 pfu. For our purposes, we are assuming all of our calculated flux reaches the upper atmosphere of the planet; for this single, energetic solar event, at least, the scaling matches well. To be sure, we must keep in mind that we are making many assumptions (e.g., M dwarf x-ray/proton scaling is similar to solar scaling, absence of Earth-like geomagnetic field). 

Here, we discuss the energy ranges and expected proton fluences for the present work's simualted flares. For the low end of our M dwarf flare energy scale, the 10$^{30.5}$ erg flare gives a $\geq$10 MeV proton flux of $\sim$3.25$\times$10$^{5}$ pfu (fluence of $\sim$1.95$\times$10$^{8}$ pr cm$^{-2}$), corresponding to a 1-8 \AA~bandpass flux of 1.25$\times$10$^{-2}$ W m$^{-2}$ from our scaling method or a GOES X125 equivalent solar flare. The top end energy of 10$^{34}$ erg scales to $\sim$6.20$\times$10$^{8}$ pfu (fluence of $\sim$9.6$\times$10$^{12}$, corresponding to 9.08 W m$^{-2}$ in the 1-8 \AA~bandpass (approximately the same flux calculated in \citet{segura2010}), or a GOES X94,000 solar flare. This is certainly a regime with much uncertainty, given no such events have ever been recorded for the sun. 

Not all flares necessarily produce energetic proton events. However, with the above analysis in mind, we use the following reasoning to assume all stellar flare events modeled in the present work have a proton event associated. It has been reported that ~100\% of solar flare events of class GOES X2 or higher have energetic proton events (e.g., CME, SEP) associated with them \citep[e.g.][]{yashiro2006,hudson2011,dierckxsens2015}. That corresponds to an X-ray flux of ~2$\times$10$^{-4}$ W m$^{‑2}$ over the 1-8 \AA~bandpass. As mentioned above, following the assumptions made in \citet{segura2010}, the lowest energy flare in the present work would be roughly equivalent to a GOES X125 flare from the Sun --- larger than any recorded event. Using this as proxy for the M dwarf events, it is not unrealistic to assume a one-to-one flare-CME frequency correlation for our work.

Since not all CMEs are guaranteed to be directed toward an orbiting planet, we want to take into account the geometries of emitted energetic proton events, \citep[e.g.,][]{khodachenko2007,kay2016}. \citet{khodachenko2007} calculates the probability of a planet being hit by a flare related CME as:

\begin{align}
    P_{CME} = \frac{(\Delta_{CME}+\delta_{pl})\;\sin\left[(\Delta_{CME}+\delta_{pl})/2\right]}{2\pi \sin\Theta} \label{eq:proton_impact}
\end{align}

\noindent where $\Delta_{CME}$ is the angular size of the CME, $\delta_{pl}$ is the solid angle subtended by the planet, and $\Theta$ is the hemispheric latitude range of CME activity on the star. Here, we study two sets of values of $\Delta_{CME}=5\pi/183$ ($\pi/2$) as a restrictive (permissive) estimate, and stellar magnetic latitude $\Theta=\pm\pi/4$ in both cases; $\delta_{pl} \approx 10^{-8}$ is negligible for an Earth-like planet at 0.16 AU; further, we assume zero orbital inclination. We also assume that CME activity expands similarly to solar geometry, instead of being trapped in the astrospheric current sheet. This gives a probability for every CME to hit our simulated planet of $P_{CME}$=0.083 (0.25) for the restrictive (permissive) CME angular size. In this work, we assume all particles in our calculated fluence impact and precipitate into the atmosphere of the planet, ignoring potential deflection from a planetary magnetic field. In \S~\ref{sec:unmagnetized}, we discuss the potential complications, and suggest directions for future work.

\subsection{Optical-red flare continuum}
\label{sec:conundruum}

In \citet{segura2010}, the flare spectrum used to drive the photochemistry included flux from the far UV to ~4500 \AA. During the 20 timesteps in the flare, flux from wavelengths greater than 4500 \AA~was not scaled to flare levels from the blackbody continuum associated with quiescence. In the present work, we extended all flare timesteps to include increased flux out to the near IR, to 8500 \AA. 

To use a more realistic spectrum, the optical-red continuum wavelengths longward of 4500 \AA~for each timestep were scaled to the relative flux value at 4500 \AA, and then extended to 8500 \AA~by adding the modeled combined tails of two blackbody continuum curves with maximum temperatures of 5,000K and 10,000K to represent the increased optical-red spectrum from the flare – the so-called ``conundruum" – following \citet{kowalski2013}. Fig.~\ref{fig:photorise} shows the extended flare spectra during the impulsive, rising phase of the flare from 2000 - 8500 \AA, with our extended, upscaled spectrum to the right of the vertical dashed red line at 4500 \AA. The timesteps in the decay phase of the flare events were similarly scaled.

\section{Results}
\label{sec:results}

The following subsections demonstrate the resulting effects of repeated stellar flares, with and without CME/SEP events, on the atmosphere and surface UV flux of an unmagnetized Earth-like planet at 0.16 AU.

The 1D photochemical model runs for each timestep throughout every simulation. For results in \S~\ref{sec:energy_param} (energy parameter-space comparison) and~\ref{sec:time_param} (temporal parameter-space comparison), the timesteps during the flare are set by the energy scaling characteristics and light curve timing from GJ1243 \citep[e.g.][]{hawley2014,davenport2014}, discussed above in \S~\ref{sec:kepler_multi}, and vary from order 10$^0$ to 10$^3$ seconds. For \S~\ref{sec:multi_emonly} (EM-only FFD generated events) and~\ref{sec:multi_emprot} (EM + proton FFD generated events), the entire flaring period for all durations were run with a 1 minute cadence; the coupled 1D radiative-convective model is run between flares, typically for intervals of order $\geq$ 10$^3$ s both during interflare periods, and the post-flare recovery phase is run with a variable timestep, i.e., the method used in \citet{segura2010}. It is worthwhile to note here that due to the high frequency of flare activity, the atmosphere does not typically return to an equilibrated steady state between flare events, as would be the case for a planet orbiting such an active stellar host. The timescale to equilibrium for individual flare events can be seen in Fig.~\ref{fig:parameter_energy}.

\subsection{Energy parameter-space comparison with \citep{segura2010}}
\label{sec:energy_param}

We simulated the atmospheric effects of an array of single-flare energies from the range covered by the GJ1243 FFD, and compared the results to the original ADLeo sized flare (10$^{34}$ ergs) simulated in \citet{segura2010}. The effects of electromagnetic-only (EM-only) and electromagnetic with proton events (EM+protons) were both simulated; the proton flux and fluence for each flare (see Fig.~\ref{fig:fluence_vs_flux}) are correlated with the amplitude of the flare as given by the Kepler observational results as described in \S~\ref{sec:kepler_multi}, Eq.~ \ref{eq:amplitude}, and as discussed above in \S~\ref{sec:proton_events}. 

Fig.~\ref{fig:parameter_energy} shows the single-flare effects on the ozone-column for EM-only and EM+protons in the left and right panels, respectively, as a function of time.  The 10$^{34}$ erg EM-only case in the left panel of Fig.~\ref{fig:parameter_energy}  is a direct comparison of the new model (dash-dotted blue line, with updates included from \S~\ref{sec:segura_ext}) to the original flare simulated in \citet{segura2010} (dashed black line). In the new treatment, the magnitude of O$_3$ column depletion is reduced, and the depressed column density recovers more gradually due to the long decay phase of the flare event. Applying the flare template from \citet{davenport2014} to the flare spectral evolution discussed in \S~\ref{sec:kepler_multi} (e.g., new treatment has wider FWHM, less impulsive, longer decay phase) is indicated in the overall impact on the O$_3$ column as seen in these single events. The results suggest that EM-only flare impulsivity has a moderate impact on the O$_3$ column, when compared to the overall flux delivered to the atmosphere, whereas the atmospheric recovery time is more dependent on the flare duration or overall energy.

For EM-only lower energy flare events ($\le$ 10$^{32}$ erg), the overall effect is to initially decrease the O$_3$ column by a small amount (factor of 10$^{-4}$), and subsequently increase the ozone column by a fractional percentage ($< 1\%$). At lower flare energy, the flare duration is fairly short (e.g., 600 seconds for 10$^{30.5}$ erg). The generation of ozone via free O atoms from photodissociation of other species (e.g., H$_2$O) occurs more rapidly than the direct photodissociation of O$_3$, initially resulting in the overall slight increase of the O$_3$ column. Once the energy of the modeled EM-only flares becomes sufficiently large ($>\sim10^{32}$ erg), the overall O$_3$ photodissociation rates are significant enough to produce a notable net loss of ozone due to the flux input from a single flare event. The long-term effects of this slight increase in O$_3$ column depth by these low energy, high frequency EM-only events is discussed and shown below in \S~\ref{sec:multi_emonly}.
 
Fig.~\ref{fig:parameter_energy}, right panel, shows the EM+protons cases, including a direct comparison of the new model (colored lines) with the results of \citet{segura2010} (dashed black line). The most significant difference featured in runs from the present work is the rapid increase and subsequent decrease in the ozone column at the time of NO$_x$ injection, and is driven by the mass-balancing addition of O at an equivalent rate to the NO injected as discussed in \S~\ref{sec:segura_unfixed}. In these events, NO$_x$ is injected at the peak of the flare, as in \citet{segura2010}, the timing of which increases with increasing overall flare energy, due to the related increase in both flare rise/decay times relative to flare durations \citep[e.g.][]{hawley2014}, as noted in Eq.~\ref{eq:duration}. The immediate subsequent decrease of the O$_3$ generated by the injected O atoms is seen, e.g., due to the rapid action of NO$_x$ destruction of O$_3$.

Comparing the two 10$^{34}$ erg flares in panel (b) (present work - blue dash dotted line, \citet{segura2010} - black dashed line) we find a difference between the two events, with a peak ozone loss of $\sim$90 (94)\% occurring $\sim$2 years after the event for the present work (\citet{segura2010}), and recovery taking $\sim$50 years in both cases. It is apparent that events with the highest impact on the atmosphere are those including stimulated NO$_x$-production through energetic proton flux into the upper atmosphere, as is evidenced by comparing the lowest EM+protons event at 10$^{30.5}$ erg, which shows a factor of $\sim$14 greater O$_3$ column loss ($\sim$8.4\%) when compared to the highest EM-only 10$^{34}$ erg flare in the left panel (0.6\%). The proton flux, and therefore NO$_x$ production, for all EM+protons events were scaled according to the method outlined in \S~\ref{sec:proton_events}, and seen in Fig.~\ref{fig:fluence_vs_flux}. This corresponds to a total fluence for the 10$^{30.5}$ erg event of $\sim$1.95$\times$10$^8$ protons cm$^{-2}$, or a factor of 2.03$\times$10$^{-5}$ of the simulated ADLeo event, and a factor of $\sim$1.33$\times$10$^{-2}$ of the Carrington event of 1859 \citep{rodger2008}. Taking into account this scaling for lower energy flare events, the ozone column is still significantly altered, with peak ozone column loss of $\sim$8.4\% occurring $\sim$2.5 months after the event; for our simulation of a Carrington-sized proton events, peak O$_3$ loss of $\sim$36\% equilibrium value occurs at $\sim$1.15 years after the flare.

Repeated impacts by proton events – even at the lowest energies considered - could be particularly impactful on the ozone column of the Earth-like planet, depending on orbital parameters, as well as CME frequency and geometries. The resulting effect of multiple proton events on the O$_3$ column and surface UV flux are discussed below in \S~\ref{sec:time_param},~\ref{sec:multi_emprot}, and~\ref{sec:uv_flux}.

\subsection{Temporal spacing of flare activity, with and without protons}
\label{sec:time_param}

We simulated the effect of multiple flares with varying temporal spacing to parameterize the effect of flare frequency on atmospheric evolution, notably the O$_3$ column. Two sets of simulations were performed, EM-only and EM+proton events. The EM-only simulations were performed with 1000 flares, over five different cases that vary the interflare separation periods at two hours, one day, one week, one month and one year. Note that the spacing between flare events is measured from the final timestep of the initial flare event and the first timestep of the following flare event. From Fig.~\ref{fig:parameter_energy}(a), one can see that single lower energy flares do not significantly impact the ozone column, so the results shown in Fig.~\ref{fig:davmultinoprot} to determine the effect of EM-only flare spacing were all performed with the ADLeo-like 10$^{34}$ erg flare flux to generate the worst case effect of UV on the ozone column. In \S~\ref{sec:multi_emonly} the effects of the lower energy EM-only events can be seen during the simulations generated from the GJ1243 FFD. Simple pairs of flare events were also simulated for both EM-only and EM+protons, but showed little deviation from single events and are not shown here.

Fig.~\ref{fig:davmultinoprot} shows the results for the EM-only simulations, consisting solely of multiple 10$^{34}$ erg flares. The results indicate that the electromagnetic events from active M dwarfs impact an atmosphere slowly, but perhaps over long enough time - significantly. The case with a one year interflare separation (dashed blue line) shows no appreciable change over the course of $\sim$1000 years, as the period of atmospheric recovery for a single 10$^{34}$ erg event is ~$\sim$1 year (as seen in Fig.~\ref{fig:parameter_energy}(a)); therefore, each flare is affecting a nearly re-equilibrated atmosphere, leading to seemingly small change over periods of time of order 1000 years. It should be noted here that the active star used in the flare model - the ADLeo spectra applied to the atmosphere via the GJ1243 FFD and light curve - would experience a flare of this magnitude approximately once every 489 days, as seen in the stellar FFD in Fig.~\ref{fig:ffdamp}. 

The cases run with interflare separation of one day, one week, and one month all suggest that extended periods of frequent events equivalent to the great flare of AD Leonis slowly erode the O$_3$ column over extended periods, but seem to reach a potential new equilibrium at $\sim$97\% of the steady state value. This gives rise to some concern for a potentially habitable planet as M dwarfs stay active for well into their multi-Gyr lifetimes on the main sequence \cite[e.g.][]{silvestri2005}. However, instead of focusing on the loss rate for those flaring frequencies, we will comment on the simulation with two hour interflare separation - the worst case scenario from a star much more active than ADLeo or GJ1243. In this case, 10$^4$ flares were run instead of 10$^3$, which corresponds to a period of $\sim$6.9 years. At that point, the O$_3$ column has been eroded by  only $\sim$7.4\% of the equilibrium value, albeit at a very steep rate of decline. 

To estimate the continued loss rate if this high amount of flux continually impacts the planetary atmosphere, we extend the dot-dashed red trendline in Fig.~\ref{fig:davmultinoprot}, and find that even by the age of the current universe ($\sim$4.3$\times$10$^{17}$ seconds) the O$_3$ column loss would only be $\approx$86.1\%. This is less impactful than the full impact of a single proton event from a 10$^{34}$ erg flare. This result is discussed primarily as an extreme thought experiment, as such conditions are highly unlikely to persist for a significant portion of planetary evolution: no M dwarf has been observed experiencing 12 AD Leonis great flare-sized events per day. It is possible that very young stars could exhibit such activity, though, so these results could be applied to very early evolution of an Earth-like atmosphere containing significant oxygen content \citep[e.g.,][]{lugerbarnes2015,meadows2017a}.

The EM+protons simulations were performed with 100 flares - all events containing proton fluence impacting the planetary system - for the same separation periods mentioned above. We chose to simulate energies of 10$^{30.5}$ erg, the lowest energy included in this work, 10$^{31.9}$ erg events with approximately equivalent proton fluence to the Carrington event, and 10$^{34}$ erg, equivalent to the ADLeo flare. In Fig~\ref{fig:davmultiprot}, we show the results from 100 proton events for three different energies: 10$^{30.5}$ - the lowest energy considered in the present work, 10$^{31.9}$ -  events with proton fluence similar to the Carrington event, and 10$^{34}$ - events with proton fluence similar to the great flare of AD Leonis simulated in Fig.~\ref{fig:parameter_energy}, respectively. The black dashed line in each panel shows the impact of a single proton event of each energy for reference. In these simulations, every event was assumed to directly impact the planet and fully affect the atmosphere. In reality, CME and SEP events could be glancing blows, or miss the planet entirely, as discussed in \S~\ref{sec:proton_events}. The interflare separations in each case are the same as those used in the EM-only simulations discussed above. 

In general, the results suggest that even the lowest energy, most frequent flares from a star like GJ1243 - 10$^{30.5}$ erg - can rapidly erode the O$_3$ column of an Earth-like planet for all interflare frequencies except the once-per-year case. Table~\ref{tab:o3_fits} shows the time to destroy $\geq$~99\% of the O$_3$ column for the cases in the middle (Carrington-like, 10$^{31.9}$ erg) and bottom (ADLeo-like, 10$^{34}$ erg) panels Fig.~\ref{fig:davmultiprot}. To predict these time to loss values, we used a log fit to the O$_3$ column depth curve for the last $\approx$1000 points during the period of active flaring, and then extrapolated to find the intercept at 99\%; the method assumes that the loss continues at the rate of the last 1000 or so points of O$_3$ column evolution during flare activity. Note that assuming the loss rate is maintained is possibly inaccurate, as one can see from the cases simulated for 10 year duration in Fig.~\ref{fig:realmultiprot} and discussed in \S~\ref{sec:multi_emprot} --- the response at long times tend to follow a separate fit, which declines more gradually. For this reason, we do not extrapolate from the 10$^{30.5}$ erg results in the top panel of Fig.~\ref{fig:davmultiprot}, instead we discuss the impact of these lower energy flares below, in \S~\ref{sec:multi_emprot}.

If every flare from a star as active as GJ1243 hits a planet, we might expect the rate of O$_3$ depletion to be quite rapid, similar to what is seen in the top panel of Fig.~\ref{fig:davmultiprot}, for the two hour, one day, and one week interflare separations. One can see that for the one year interflare separation, the trend flattens out and the atmosphere reaches a new steady state at $\sim$86\% of the original steady state column depth. Similarly, for the one month interflare simulations, the trend begins a flattening just before the end of the flare activity, likely reaching a new equilibrium around 50\% of steady state. 

Taking into account the CME/SEP geometries discussed in \S~\ref{sec:proton_events}, along with the FFD, the most likely case for a planet orbiting a star like GJ1243 to experience a Carrington-like event is in bold in Tab.~\ref{tab:o3_fits}. For the Carrington-like 10$^{31.9}$ erg flares, one flare occurs every $\sim$3.7 days. Assuming the CME probability of 0.08, this implies one proton event hits the planet every $\sim$46 days. In this case, applying the power law fit suggests an O$_3$ lifetime of $\sim$2.64$\times$10$^{14}$ seconds, or $\sim$8.4 Myr for a planet's O$_3$ column to be effectively destroyed with respect to surface UV shielding. Note that prolonged simulations may show a change in the trend as the atmosphere is able to recover from multiple flaring events such as these -- see discussion in \S~\ref{sec:multi_emprot}.

In the case of the multiple 10$^{34}$ erg flares, one can see in the bottom panel of Fig.~\ref{fig:davmultiprot} and in Tab.~\ref{tab:o3_fits} that each interflare frequency simulated has the potential to effectively destroy the O$_3$ column to less than 99\% in less than 3 years for all cases --- except the two hour and one year interflare separations. While proton events impacting the planet with this frequency are unlikely for a host like GJ1243, a more active M dwarf, or very active early star could produce such events. A host like GJ1243 would experience one 10$^{34}$ erg flare once every $\sim$489 days, so the once per year case (blue dashed line) is rough approximation to the expected atmospheric response if each of the events hit the planet. This case indicates that the O$_3$ column is roughly equilibrated at $\sim$94\% loss. However, as the discussion in \S~\ref{sec:proton_events} notes, approximately one in eight (or one in four) proton events of all energies hits the planet, depending on CME geometry.

For all of these EM+proton simulations, the results in Tab.~\ref{tab:o3_fits} assume that only these proton events are impacting the planetary atmosphere, ignoring other less and more energetic proton events ejected by the stellar host as would be seen in a realistic flare distribution such as Fig.~\ref{fig:dav_ffdflares}. The simulations including realistic flare distributions and proton event geometries are shown and discussed below in \S~\ref{sec:multi_emprot}.

\subsection{EM-only FFD-generated flares} 
\label{sec:multi_emonly}

The results of \citet{segura2010} and those in \S~\ref{sec:energy_param} above indicate that single, EM-only flares do not significantly impact the atmosphere of the ozone column of an Earth-like planet at 0.16 AU. Results in \S~\ref{sec:time_param} suggest that even frequent 10$^{34}$ erg EM-only events are not as impactful as a single, lower-energy proton event. However, given the flaring frequency of active M dwarfs like GJ1243 \citep{hawley2014}, terrestrial planetary atmospheres for planets orbiting these active hosts will be impacted by multiple flares per day at varying energies. Fig.~\ref{fig:ffdamp} shows the GJ1243 FFD and amplitude for the stellar host's flare activity in our simulations, indicating that $\sim$7 flares per day of energy 10$^{30.5}$ erg and above will impact the planetary atmosphere.

\begin{figure}
\centering
\includegraphics[width=0.75\textwidth]{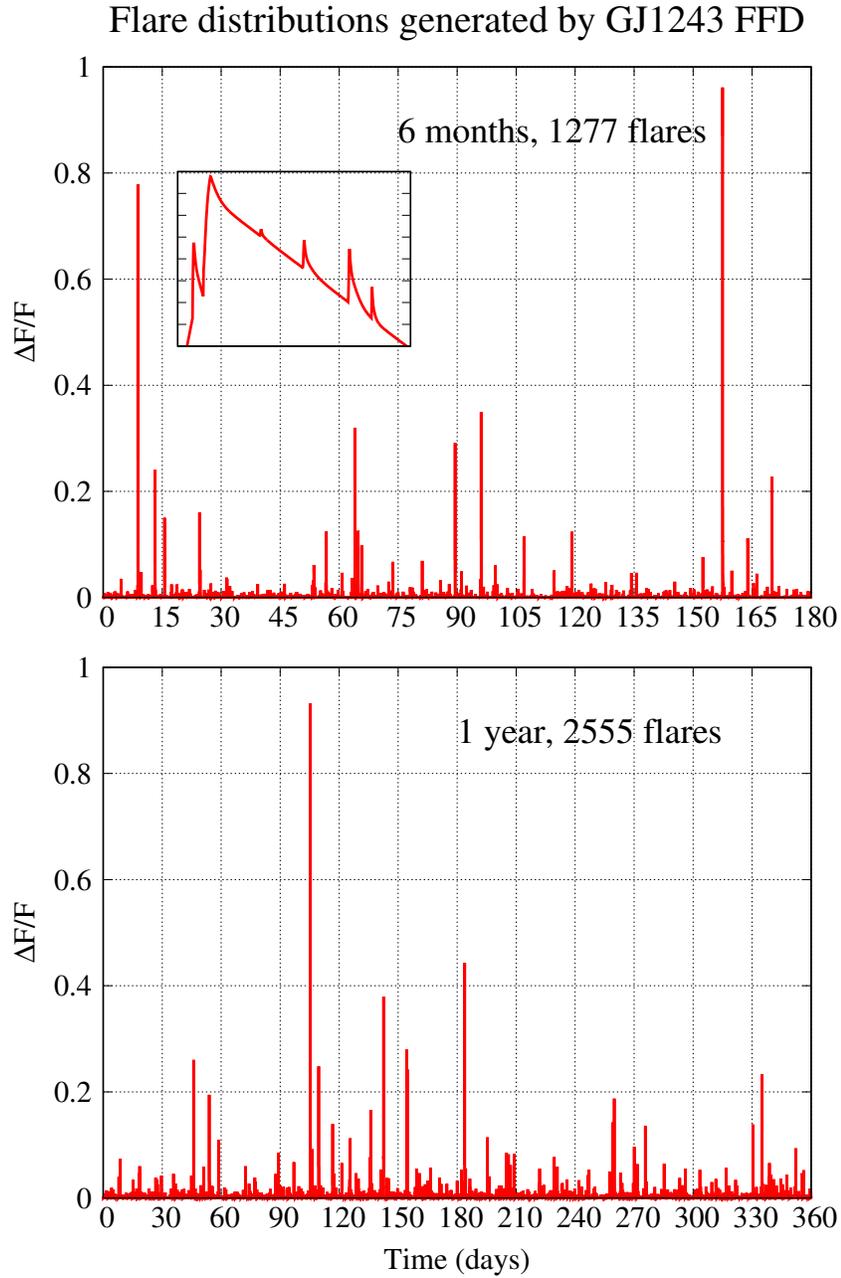}
\caption{{\bf Top:} Six months, flare distribution (1277 events) generated from the GJ1243 FFD. The inset identifies an example complex flare near $\sim$9 days, produced by flare flux stacking. {\bf Bottom:} One year flare distribution (2555 events) from the GJ1243 FFD.}
\label{fig:dav_ffdflares}
\end{figure}

\begin{figure}
\centering
\includegraphics[width=0.75\textwidth]{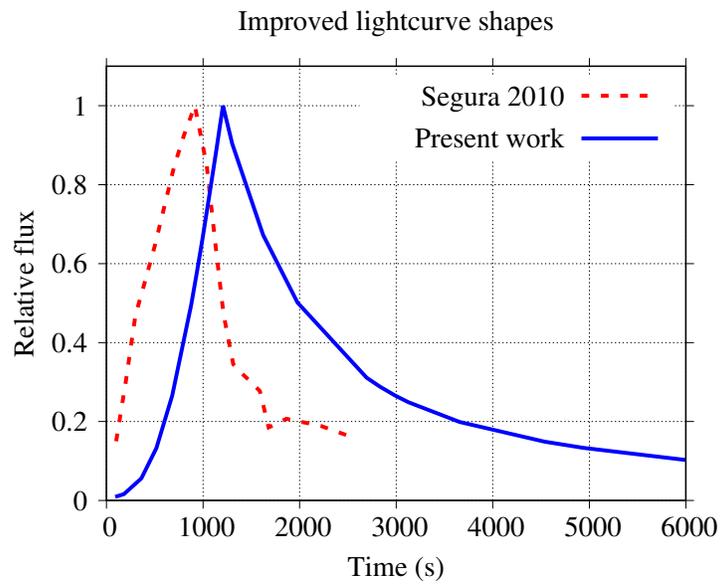}
\caption{Comparison of light curve evolution from \citet{segura2010} vs the method adopted in the present work, based on the empirical modeling results of \citet{davenport2014}.}
\label{fig:lightcurves_comp}
\end{figure}

\begin{figure}
\centering
\includegraphics[width=0.75\textwidth]{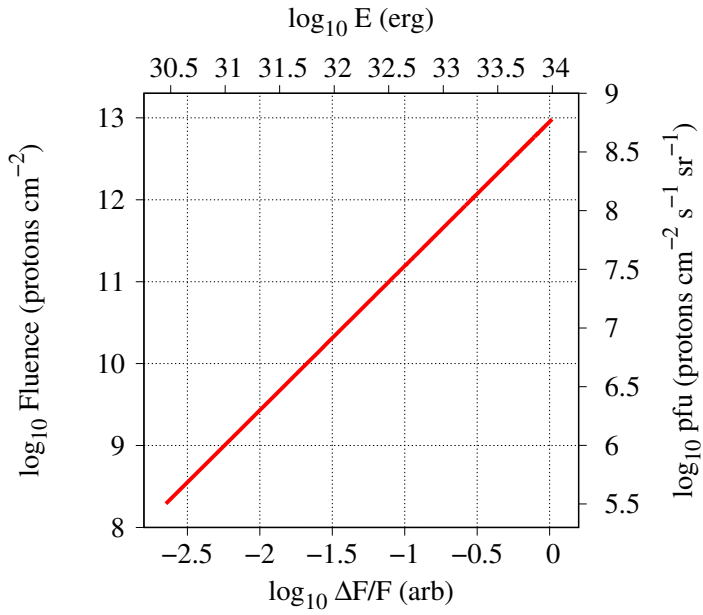}
\caption{The relationship of proton fluence (left y-axis) and proton flux units (pfu, right y-axis) for the flares simulated in the present work, as function of relative flux increase (bottom x-axis) and total flare energy (top x-axis).}
\label{fig:fluence_vs_flux}
\end{figure}

\begin{figure}[!h]
\centering
\includegraphics[width=0.75\textwidth]{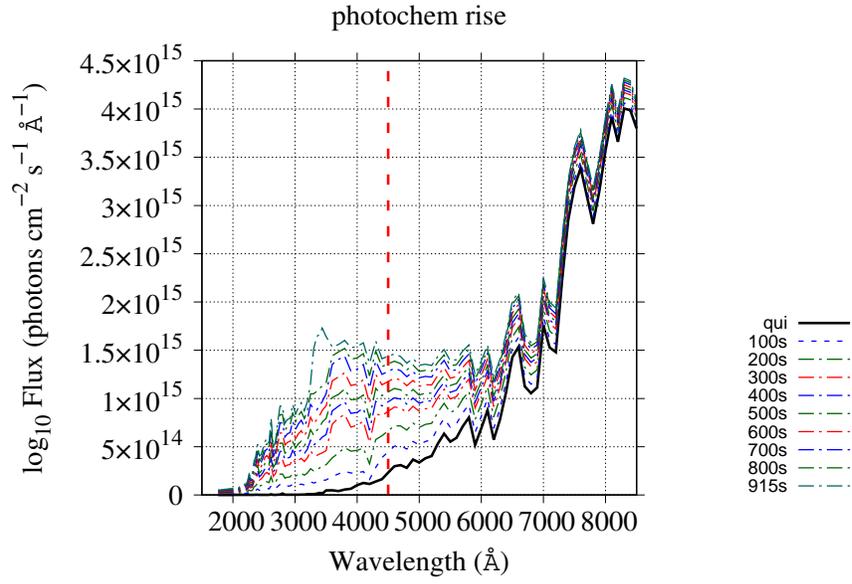}
\caption{Full UV-visible spectrum for the impulsive phase of an AD Leonis great flare sized (10$^{34}$ erg) event. The spectra to the right of the vertical dashed red line have been added in the present work.}
\label{fig:photorise}
\end{figure}

\begin{figure}
\centering
\includegraphics[width=0.9\textwidth]{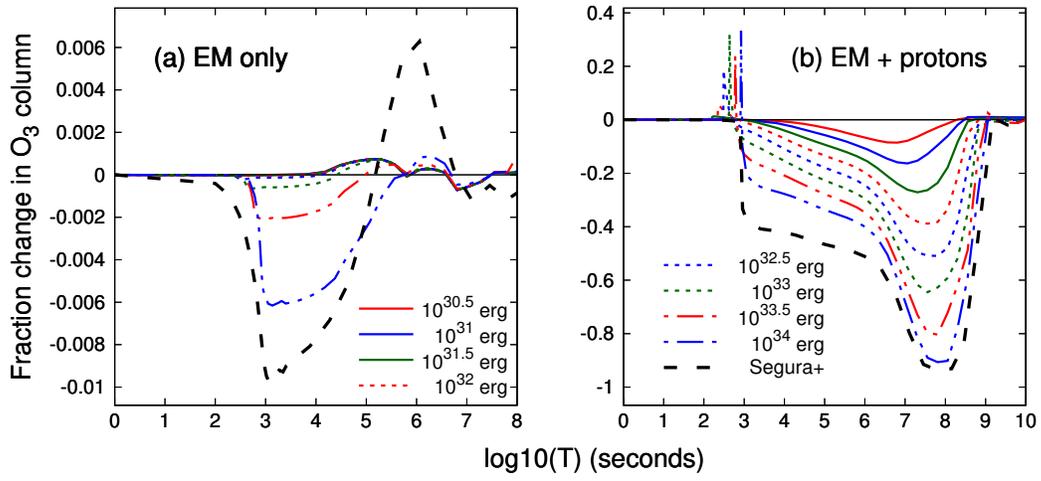}
\caption{ O$_3$ evolution for single-flare energy parameter comparison with \citet{segura2010} for EM-only (left panel) and EM+protons (right panel) 10$^{30.5}$ - 10$^{34}$ erg flares.}
\label{fig:parameter_energy}
\end{figure}

\begin{figure}
\centering
\includegraphics[width=0.75\textwidth]{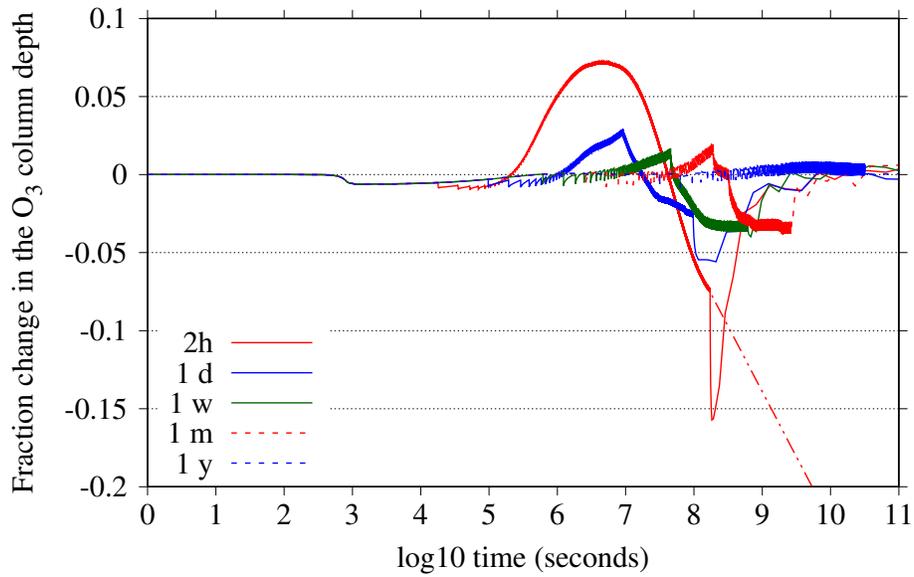}
\caption{The effects on the O$_3$ column of EM-only, 10$^{34}$ erg flares with varying interflare separations. Separations of 1 day, 1 week, 1 month and 1 year included simualtions of 10$^3$ flares, where the 2 hour separation included 10$^4$ flares to obtain extended effects for long-term prediction of O$_3$ column (dash-dotted red line).}
\label{fig:davmultinoprot}
\end{figure}

\begin{figure}
\centering
\includegraphics[width=0.5\textwidth]{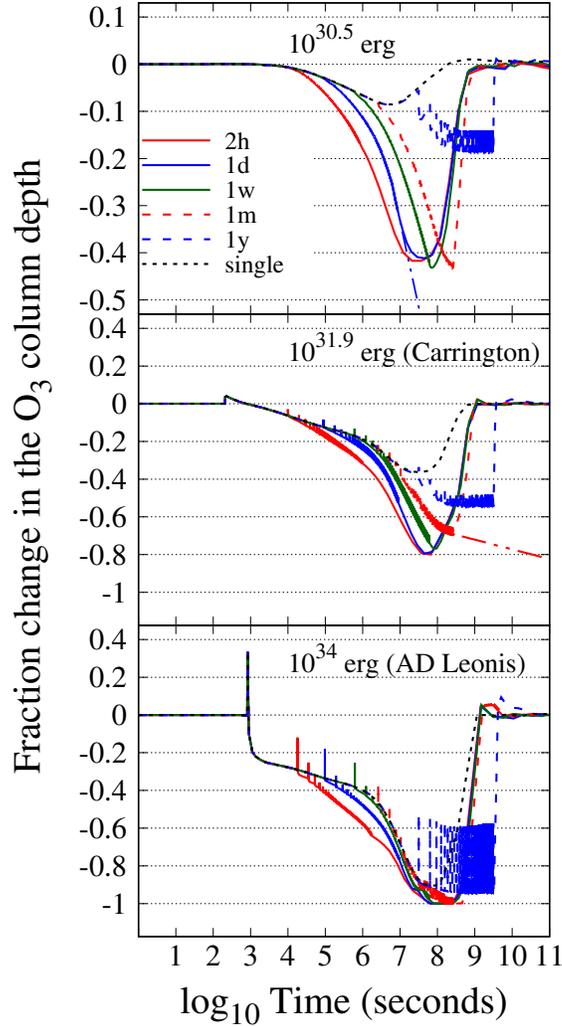}
\caption{O$_3$ evolution driven by repeated proton events, 100 flare events for all cases. {\bf Top:} 10$^{30.5}$ erg flares+protons. The 1 d$^{-1}$ period is likely to occur for a planet orbiting GJ1243; the dash-dotted blue line extrapolates the predicted O$_3$ loss rate. {\bf Middle:} Carrington equivalent proton events at 10$^{31.9}$ erg. The dash-dotted red line predicts O$_3$ evolution for the most likely frequency experienced at a GJ1243 orbiting planet. {\bf Bottom:} AD Leonis equivalent proton events with 10$^{34}$ erg.}
\label{fig:davmultiprot}
\end{figure}

\begin{figure}
\centering
\includegraphics[width=0.8\textwidth]{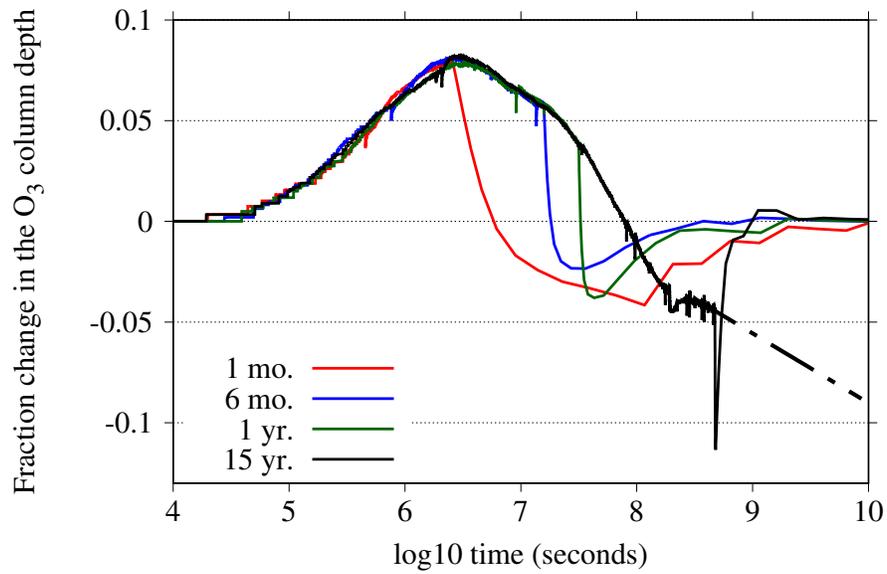}
\caption{O$_3$ evolution for EM-only flare events generated from the GJ1243 FFD for periods of 1 month, 6 months, 1 year, and 15 years. The dash-dotted black line predicts continued effects of flaring beyond 15 years.}
\label{fig:realmultinoprot}
\end{figure}

A six month (one year) example distribution of the generated flares are shown in the top (bottom) panel of Fig.~\ref{fig:dav_ffdflares}, showing a generated timeline of 1277 (2555) flares over that 180 (360) day period. The flux for each flare was stacked additively, allowing for the effects of simulated complex flares as shown in the log-scale inset for the top panel of Fig.\ref{fig:dav_ffdflares}; the inset is a zoom around the large flare event of amplitude $\sim$0.78 ($\sim$10$^{33.84}$ erg) just prior to day 9. Four such individually generated distributions were created using the GJ1243 FFD and used to drive a more realistic atmospheric simulation response to EM-only flaring for periods of: one month, six months, one year, and 15 years. In the six month (one year) simulations shown in Fig.~\ref{fig:dav_ffdflares}, the largest flare is of amplitude $\sim$0.972 (0.950) or 9.6(9.3)$\times$10$^{33}$ erg. An average of seven flares occur per day in these distributions, most of low energy, and are not visible due to the scaling of the figure. The effects on the ozone column of the planet from the EM-only, multiple flaring distributions are shown in Fig.~\ref{fig:realmultinoprot}. 

The simulations for each of the four distributions show that the early ozone evolution is dominated by the multitude of smaller flares. These smaller events increase the overall ozone column steadily and in predictable fashion despite each run being an independent distribution, most easily seen as the varied presence of larger flare events throughout the flaring period, as seen in Fig.~\ref{fig:dav_ffdflares}. The larger flare events ($>~10^{33}$ erg) can be seen as negative spikes throughout the distributions. All runs reach a peak ozone increase of ~8\% around one month, and the longer runs then turn back toward equilibrium. The 15 year case runs long enough to drive the ozone column into the loss region, and from the trend for the range of 2-3$\times$10$^8$ s, one can extend the rate of change to calculate that this level of maintained loss would result in O$_3$ column loss of $\sim$37\% by the age of the universe; compared to the results shown in Fig.~\ref{fig:davmultinoprot}, the two hour interflare period for multiple 10$^{34}$ erg flares drove more than twice the O$_3$ depletion in the same period. It is likely that the EM-only flaring can cause a slow change over 10s of billions of years, but even small, continuous proton events combined with the EM are significantly more destructive with respect to O$_3$ abundance.

Longer periods still need to be simulated to isolate the longer term effects of EM-only flaring, and are planned for future work.  Given that M dwarf stars are active for significant portions of their long lives, these results suggest that the ozone column around these stellar hosts are likely to be depleted, quite significantly if proton events are involved, as shown in the next section.

\subsection{EM+protons FFD-generated flares}
\label{sec:multi_emprot}

To simulate a more realistic planetary atmospheric response to proton events from a stellar host with flare properties similar to GJ1243, we simulated several cases over one year of realistic flare activity generated from the FFD, shown in the bottom panel of Fig.~\ref{fig:dav_ffdflares}. The 10 year simulations were run from a flare distribution (not shown) generated independently from the GJ1243 FFD, and the distribution is separate from the six month and one year examples shown in Fig.~\ref{fig:dav_ffdflares}. To gauge the effect of CME geometry discussed in \S~\ref{sec:proton_events} and noted in Eq.~\ref{eq:proton_impact}, we ran two cases - one conservative case, with smaller CME solid angles ($\Delta_{CME}$ = $5\pi$/18) which gave each CME an $\sim$8.3\% chance to hit the planet, and one more permissive case with larger CME solid angles ($\Delta_{CME}$ = $\pi$/2), giving each CME a $\sim$25\% chance to hit. In both cases the maximum stellar latitude of magnetic activity driving proton event ejections was constrained to be between $\Theta$ = $\pm\pi$/4. With the combination of these two tools - FFD flare distribution from observation of GJ1243 and CME geometry - these results represent a more realistic representation of the effects of a proton event on the atmospheric evolution for an unmagnetized, Earth-like planet in the habitable zone at 0.16 AU around an active M dwarf host.

\begin{figure}
\centering
\includegraphics[width=0.8\textwidth]{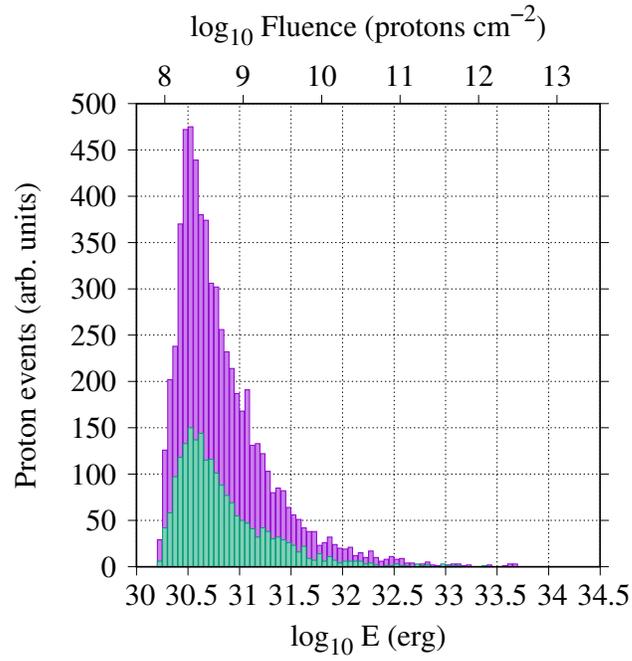}
\caption{Distribution of proton events by flare event energy and proton fluence for the 10 year simulations. CME probability P = 0.25 (0.08) is shown in magenta (green).}
\label{fig:energy_fluence}
\end{figure}

A representative frequency distribution of proton events impacting the planet is shown in Fig.~\ref{fig:energy_fluence}; the events were taken from the 10 year duration simulation, the results of which are plotted in Fig.~\ref{fig:realmultiprot} and discussed below. In Fig.~\ref{fig:energy_fluence}, the events plotted in green (magenta) represent the CME impact probability of P$_{CME}$=8\% (P$_{CME}$=25\%), as discussed above. Note there are a few events with less energy and fluence than 10$^{30.5}$ erg shown here, as proton events were checked against P$_{CME}$, and simulated at one of three timesteps during each flare event in the simulation: either the peak of the flare, or the timestep to either side of the flare peak, which allowed the proton fluence to be scaled slightly downward for some events, e.g., to 10$^{30.2}$ erg equivalent fluence. The proton events in each case were selected randomly in real-time during each simulation, and so the overall distribution of event energies and impact frequencies vary slightly in each of the four, one year simulations (the randomly selected distribution for the 10 year cases can be seen above in Fig.~\ref{fig:energy_fluence}).

The influence on the O$_3$ column is shown in Fig.~\ref{fig:realmultiprot}, with the conservative results ($\Delta_{CME}$ = $5\pi/18$, P$_{CME}$ = 8\%) in the top panel, and the more permitted results ($\Delta_{CME}$=$\pi/2$, P$_{CME}$ = 25\%) in the lower panel. The vertical dashed lines correspond to the end of the flaring periods, i.e., one and 10 years. The number of impacting proton events in the conservative (more permissive) case in the top (bottom) panel was 199.25$\pm$17.23 (595.25$\pm$11.59) out of 2555 flares in the one year distribution (see bottom panel of Fig.~\ref{fig:dav_ffdflares}), and 2045 (6301) events out of 25,550 impacted the atmosphere of the planet for the 10 year simulations, for P$_{CME}$=8\% (25\%).

The O$_3$ column responded similarly in each of the four one-year cases (shown as the average - black solid line, and the standard deviation - violet shaded region), for both restrictive and permissive CME geometries. There deviation seen in the one year simulations is due to the varying energies from the randomly selected impacting CME events (these can be seen as sudden increase of O$_3$ in each case similar to that discussed in \$~\ref{sec:energy_param}). The more conservative CME geometry resulted in $\sim$74.10$\pm$2.54\% O$_3$ column loss at the end of the year long simulations, and the more permissive geometry resulted in $\sim$80.07$\pm$4.03\% loss. 

To determine a more long term effect, one 10 year run was performed for each CME impact probability, and these are shown in red in both the top and bottom panels of Fig.~\ref{fig:realmultiprot}. The 10 year flare distribution was generated separately from the one year distribution, and the proton event selection was random and could occur on either the timestep of the flare peak, or one timestep to either side. The O$_3$ column on the last timestep for the P$_{CME}$=0.08 (0.25) CME impact case is $\sim$10.2\% ($\sim$8\%) of the steady-state equilibrium value. However, note that at the end of these simulations, both cases show a continuing trend of increasing loss. The P$_{CME}$=0.25 case, in particular, shows a consitent slope that predicts the time to 99\% (99.9\%) O$_3$ loss as 3.98$\times$10$^{12}$ s, or 126.7 kyr (1.59$\times$10$^{13}$, or 502.9 kyr), assuming the same loss rate continues.

It is difficult to simulate the behavior at longer timescales, due to the balance maintained between the timestep granularity with which these events need to be simulated for accuracy, and the exponential timescales on which these processes occur, and could feasibly impact habitability.

\section{Discussion}

All of the results in the present work have been performed with a beginning point from an equilibrium Earth-like atmosphere, which by definition assumes an oxygen-rich state. This is worth keeping in mind during the below points of discussion. Future work is planned to investigate varying steady-state conditions for potentially habitable planets in different phases of atmospheric evolution, e.g., atmospheres with high CO$_2$, low O$_2$, or haze layers.

\subsection{Multiple events, event frequency, stellar activity}
\label{sec:multiple_events}

Below a certain level of proton fluence, the response time of the atmosphere is sufficiently rapid to slow the rate of O$_3$ loss, or reach a new equilibrium. This effect can be seen in Fig.~\ref{fig:davmultiprot} as the flattening of the O$_3$ column response towards the later end of the simulations with larger interflare spacing. However, even for smaller flare energies with sufficient frequency - and therefore sustained proton flux - is it possible that O$_3$ recovery is insufficient to prevent a rate of sustained loss to levels $\leq$0.99\% of steady-state equilibrium value. 

\begin{figure}
\centering
\includegraphics[width=0.95\textwidth]{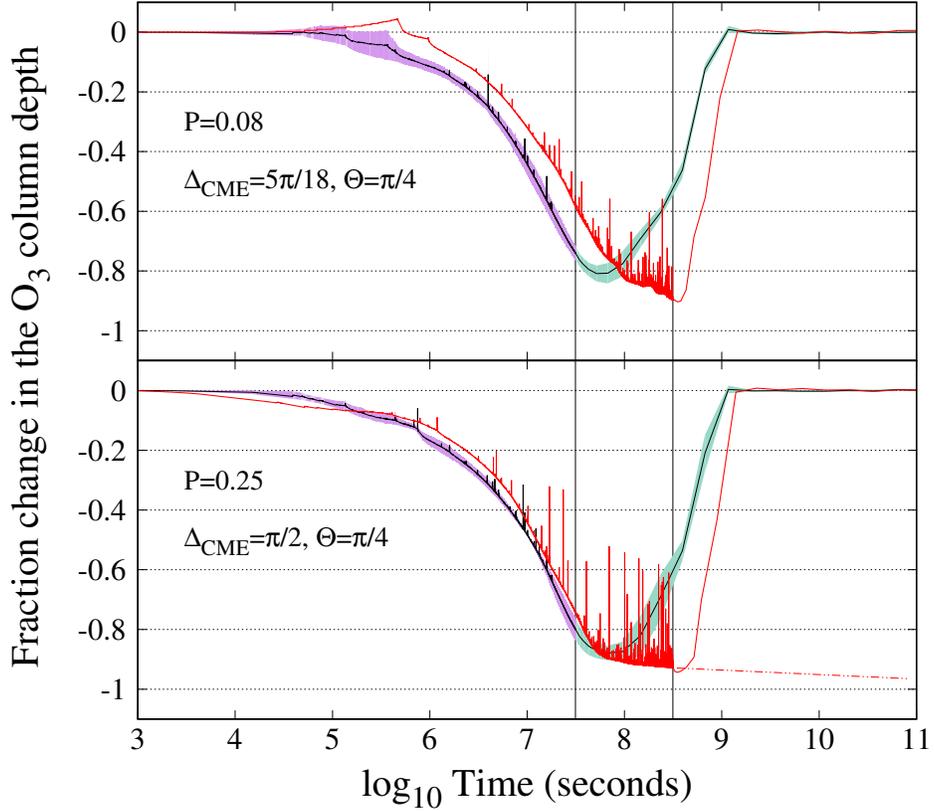}
\caption{O$_3$ column depth response multiple proton events, generated by the GJ1243 FFD and taking into account CME geometries. The average (black line) and standard deviation (shaded violet/green regions) for one-year simulations, and one 10-year simulation (red line) are shown. {\bf Top:} Events with more conservative CMEwith per-event probability for impact of P=0.083. {\bf Bottom:} Events with more-permissive geometry with per-event impact probability of 0.25.}
\label{fig:realmultiprot}
\end{figure}

The results for 100 10$^{34}$ erg flare events at a frequency of once per year in the bottom panel of Fig.~\ref{fig:davmultiprot} show the O$_3$ column reaching equilibrium at a loss of $\sim$89\% of the steady-state column depth. This outcome represents the approximate frequency of flaring at this energy level from an FFD like that of GJ 1243 (one 10$^{34}$ erg event every $\sim$489 days), and is a loss of UV shielding similar to 
\citet{segura2010} (which is important due to flaring with a depressed O$_3$ column as discussed below in \S~\ref{sec:uv_flux}), but doesn't take into account the multiple smaller events, which as discussed in \S~\ref{sec:multi_emprot} are very impactful. The results of the more realistic case in Fig.~\ref{fig:realmultiprot}, which is based on the FFD of GJ 1243, show a non-equilibrium, continual decrease in O$_3$ column depth to $\sim$94\% O$_3$ loss after only 10 years, an order of magnitude smaller timeframe than the artificial 100 flares in 100 years above. This implies the continued, rapid deposition of proton events at the low end of the FFD energy range ($\sim$7 per day for energies $\geq$10$^{30.5}$ erg) contributes significantly to the depletion of the O$_3$ column, despite a fluence that is less by a factor of $\sim$4$\times$10$^4$ when comparing the two endpoint values in our adopted FFD (10$^{30.5}$ and 10$^{34}$ erg); the frequency of the lower energy events is approximately three orders of magnitude higher. 

This brings up the issue of whether proton event frequency or fluence dominates the O$_3$ column depletion, which has implications for atmospheric response for planets orbiting smaller, older, or less active stellar hosts. In effect, the average proton fluence delivered over a period of time from GJ1243-level activity is comparable to that of a single moderate flare from the same star. For example, in Fig.~\ref{fig:realmultiprot}, the total proton fluence for the 10-year case with more restrictive CME geometry in the top panel was $\sim$9.9$\times$10$^{12}$ protons cm$^{-2}$, roughly equivalent to one 10$^{32.7}$ erg flare produced proton event per year; in the bottom panel, the fluence is roughly equal to one 10$^{33.3}$ erg event per year. From Fig.~\ref{fig:davmultiprot}, we can interpolate between the Carrington-sized events in the middle panel, and the AD Leo sized events in the bottom panel. Doing so, for fluence from single events per year, the values would equilibrate between roughly 70-75\%, not the case we see in Fig.~\ref{fig:realmultiprot}, where the 10 year simulations reach O$_3$ depletion $\geq$90\%. For the delivery of proton fluence to the planet, it appears that the frequency of events is as important, if not more important, as the total fluence from the events. Note that it is likely that GJ1243, and other highly active M dwarfs, experience flare events with lower energy and higher frequency than we modeled here, but are challenging to measure due to observational constraints.

The total fluence delivered over the P=0.08 (P=0.25) single year simulations in Fig.~\ref{fig:realmultiprot} is 1.49$\pm$0.38$\times$10$^{12}$ (4.93$\pm$0.33$\times$10$^{13}$). The size of a single Carrington event in the present paper is 1.1$\sim$10$^{11}$ protons cm$^{-2}$, and occurs roughly once per 3.7 days. In comparison, for solar cycles 19-24, the total proton fluence from all measured proton events was of order 10$^{10}$-10$^{11}$ over the 11-12 year periods \citep[e.g.,][]{shea1992,mewaldt2005,lario2011}. 

Planets orbiting active M dwarf stars are likely to experience, in a matter of months, massive O$_3$ depletion from proton flux that is multiple orders of magnitude higher than what Earth experiences over entire solar cycles; for planets in the nominal HZ around these hosts, the particle flux is further increased by a smaller orbital distance compared to Earth, scaling as 1/R$^2$, assuming isotropic expansion. For more active fully-convective M3-M5 stars, or active binary systems such as GJ1245AB, one assumes the situation is even more extreme.

\subsection{Impact on surface UV flux}
\label{sec:uv_flux}

Photo-dissociating and photo-ionizing short wavelength radiation is highly relevant to organic complexity, on one hand, it may be one of the drivers of prebiotic chemistry \citep[e.g.][]{beckstead2016life,rapf2016sunlight, ranjan2017}. On the other hand, UV radiation is responsible for mutations and degradation or transformation of biomolecules which may result in the loss of biological functions \citep[e.g.][]{de2000effects}; therefore the shielding effects of an O$_3$ layer for any potentially habitable planet is beneficial to the propagation of organisms on the planetary surface. For example, in \citet{segura2010}, the authors found that the surface UVC flux significantly increased from $<10^{-14}$ up to 10$^{-5}$ W m$^{-2}$ during the peak and recovery phase after an AD Leonis-sized proton event. This is nine orders of magnitude increase over the quiescent value, and an UV dose rate for DNA damage of the order of 10\% larger of such a planet.

In the present work, we focus on two aspects: 1) the overall effect of multiple flares on the O$_3$ column - and therefore UV shielding, and 2) the intensity of UV  surface flux from flare activity on a planet with an atmosphere already depressed by multiple, prior proton events. 

Regarding, point (1) above, we have shown that for stellar hosts with activity such as GJ1243, an O$_3$ column with the starting density of Earth can be reduced by an order of magnitude in only a few years. After this initial rapid loss, the atmosphere seems to move towards a new, lower O$_3$ equilibrium, with the loss rate slowing though not completely at equilibrium. For hosts less active than GJ1243, the situation could be analogous to the top panel of Fig.~\ref{fig:davmultiprot}, the atmosphere could reach a new equilibrium in a matter of decades.

\begin{figure}
\centering
\includegraphics[width=0.95\textwidth]{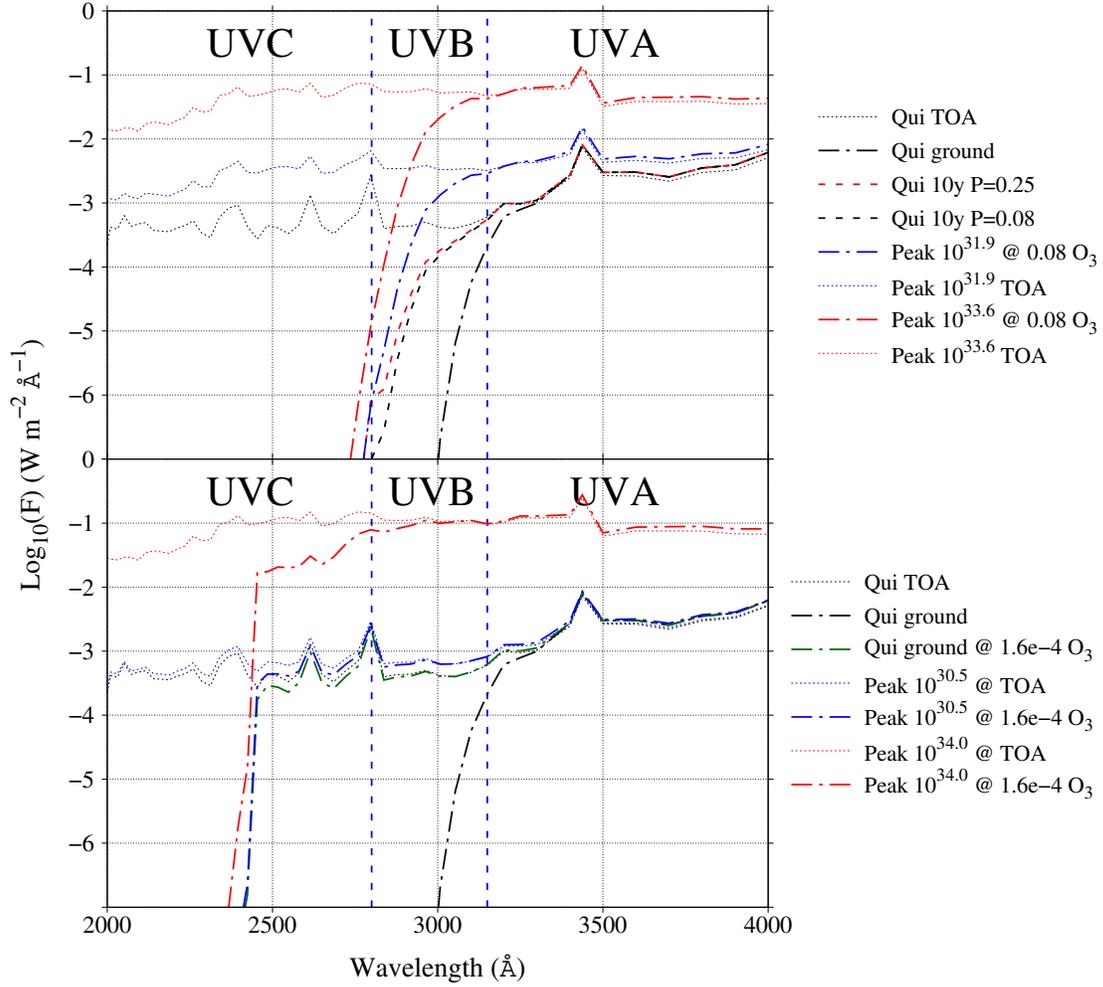}
\caption{UV flux for {\bf (top)} GJ1243 FFD generated flares, and {\bf (bottom)} extreme O$_3$ loss. Steady-state O$_3$ column at the top of the atmosphere (TOA, dotted black line) and planetary surface (dash-dotted black line); conditions with depleted O$_3$ column at surface (green dash-dotted line); at top (bottom) conditions at the peak of a 10$^{31.9}$ (10$^{30.5}$) erg flare at TOA (blue dotted line) and surface (blue dash-dotted line); conditions at the peak of a 10$^{34}$ erg flare at TOA (red dotted line) and surface (red dash-dotted line). Integrated UVC flux values are given in Table~\ref{tab:uvflux}.}
\label{fig:uvflux}
\end{figure}

Addressing point (2) above, the top panel of Fig.~\ref{fig:uvflux} represents the results of the O$_3$ depletion shown in both panels of Fig.~\ref{fig:realmultiprot}. The vertical dashed blue lines denote the limits of UVC ($<$ 2800 \AA), UVB (2800-3150 \AA) and UVC (3150-4000 \AA) regions. The quiescent UV flux is shown for initial steady-state at the top of the atmosphere (TOA, black dotted) and at the ground (black dash-dotted); also shown is the quiescent flux at the ground after the 10 year flaring period for CME impact probabilities of P=0.08 (dashed black) and P=0.25 (dashed red). UV flux for two flares near the ends of the 10 year runs are shown, at 10$^{31.9}$ and 10$^{33.6}$ erg for the P=0.25 run, with the O$_3$ column at $\sim$8\% of steady-state equilibrium value. Top of atmosphere fluxes are shown for the flares as well, with the red-dotted line representing the more energetic flare, and the blue dotted line the lower.

The O$_3$ column at the end of the proton event simulations were $\sim$0.1 and $\sim$0.08 of the equilibrium values (as discussed above in \S~\ref{sec:multi_emprot} for the P=0.08 and P=0.25 CME probabilities, respectively. Note, however, that the O$_3$ loss would likely continue with longer duration simulated flaring and proton fluence. Based on the simulations in \citet{segura2010}, one doesn't expect much UV-C flux to penetrate to the ground given those column O$_3$ abundances even at higher energies, as one can see in the surface UV for various conditions listed in Table~\ref{tab:uvflux}. During quiescence, the value of UV-B and UV-C increase substantially for all simulated cases. At the end of the two 10 year runs in Fig.~\ref{fig:realmultiprot}, the UV-C reaching the surface at the peak of representative flare events is $\sim$17-298 $\mu$W m$^{-2}$. While this $\sim$8-9 orders of magnitude over the steady-state, quiescent UV-C flux, still less than the UV-C flux calculated for Earth, 3.9 and 2 Ga ago \citep[Table 6][]{rugheimer2015uv,arney2016}. Based on Earth history this amount of UV-C radiation may not be a substantial detriment to habitability. For a Carrington-sized flare during the period when the O$_3$ column is depressed to 8\% of steady-state value, $\sim$17 $\mu$W m$^{-2}$ UV-C reaches the surface. For a near ADLeo-sized flare, 10$^{33.6}$ erg, $\sim$298 $\mu$W m$^{-2}$ reaches the surface at the peak of the flare. An ADLeo sized flare at these O$_3$ levels would peak at $\sim$740 $\mu$W m$^{-2}$.

However, as discussed in \S~\ref{sec:kepler_multi}, the decay phase of the flare dominates the energy input. If one integrates over the temporal evolution applied in the present work, the total amount of UV-C energy delivered to the surface for the 10$^{33.6}$ erg event (with 8\% equilibrium O$_3$ column) is 0.431 J m$^{-2}$ over the $\sim$10$^{4}$ s event, giving an an average UV-C flux of $\sim$43.1 $\mu$W m$^{-2}$. For the ADLeo event, this figure would be a factor of $\sim$2.5 larger, due to the increased event amplitude and duration. Obtaining a 90\% kill rate for a commonly studied, radiation resistant bacterial species, {\it Deinococcus Radiodurans}, requires a dose of $\sim$550 J m$^{-2}$ \citep{gascon1995}, which requires more than 1200 of such 10$^{33.6}$ erg flare events at an 8\% equilibrium O$_3$ level.

The average energy of all flare events for the 10 year distributions is $\sim$10$^{31.4}$ erg, each of which - when integrated - each provide a UV-C dose of only $\sim$0.13 mJ m$^{-2}$ to the planetary surface with 8\% of Earth's equilibrium O$_3$ column. One should consider the effect to any underlying precursor organic molecules for which UV light may be a driver of complexity \citep[e.g.][]{beckstead2016life,rapf2016sunlight, ranjan2017}, however, regarding habitability. Another consideration is that the O$_3$ loss in these cases is still decreasing. 

The bottom panel of Fig.~\ref{fig:uvflux} represents a worst-case scenario, where the O$_3$ column has been depleted to $\sim$10$^{-4}$ of steady-state equilibrium value. This occurs in our simulations with repeated, extreme flaring with high proton fluence. In the present work such a state is reached after only $\sim$4.9 yr, given one ADLeo sized proton event per month - as seen in the bottom panel of Fig.~\ref{fig:davmultiprot}. While such a flare frequency for that energy is not observed from GJ1243 at its current age, early stellar hosts and perhaps more highly active M dwarf hosts could exhibit such activity for millions of years. It is instructive to quantitatively consider these UV fluxes, as determining these values could offer insight to constraining the early development of complex organics or life on planetary surfaces - not the depths of any ocean, however.

Fig.~\ref{fig:uvflux}, bottom panel, shows the UV flux in W m$^{-2}$ \AA$^{-1}$ for four different cases: 1) quiescent UV flux at the top of the atmosphere (TOA, black dotted line) and at the surface (black dash-dotted line), 2) quiescent flux with the O$_3$ column depleted by a factor of 1.6$\times$10$^{-4}$ at TOA (green dotted line) and surface (green dash-dotted line), 3) the peak flux of a 10$^{30.5}$ erg flare with the highly depleted O$_3$ levels at TOA (blue dotted line) and surface (blue dash-dotted line), and 4) the peak flux of a 10$^{34}$ erg flare with the highly depleted O$_3$ levels at TOA (red dotted line) and surface (red dash-dotted line). 

The primary effect of the highly depleted O$_3$ layer is to allow more overall UV flux at the surface, but importantly the increased surface flux is in the UVB and UVC regions $\geq$2000 \AA. Even at lower O$_3$ levels, any extremely lower wavelength flux is likely to be absorbed by the H$_2$O and CO$_2$ present in the atmosphere \citep[e.g.,][]{ranjan2017}. We have integrated UVC flux for each of the examples in Fig.~\ref{fig:uvflux}, the results of which are displayed in Table~\ref{tab:uvflux}. Note that when the O$_3$ is depleted by a factor of order 10$^4$, even the quiescent, background spectrum of the M dwarf is sufficient to deliver 0.18 W m$^{-2}$ to the surface of the planet. Given the UV dose required for a germicidal dose, $\sim$10 J m$^{-2}$, this indicates that the surface would be sterilized within a few minutes, but likely before the O$_3$ column was depleted to this level. This level of flux does not take into account the likely continuation of frequent flare activity during the high O$_3$ depletion.

During the flare of lowest energy in the present work - 10$^{30.5}$ - peak UVC flux at the surface is $\sim$30\% higher that during stellar quiescence, at $\sim$0.24 W m$^{-2}$. It is important to note here, that approximately seven of these flares occur daily on an active M dwarf like GJ1243, and have duration on the order of tens of minutes. However, even without the presence of these numerous, frequent, low-energy flares, the quiescent value alone is relatively high. During the peak of a large, AD Leonis like flare, 60.8 W m$^{-2}$ bathes the planetary surface; over the course of the multi-hour flare, that value ranges from $\sim$1\% of that flux up to the peak value. In particular, the rise and decay phases of our simulated flare lasts $\sim$4 hours, delivering $\sim$127 kJ m$^{-2}$ of UVC to the surface during that time, compared to 35.3 J m$^{-2}$ for the smaller 10$^{30.5}$ erg flares that occur several times daily. In these cases, the survival of even the hardiest of known bacterial species is in question.
 
As pointed out by \citet{ranjan2017}, laboratory experiments are needed to evaluate how UV fluxes like the ones expected at planets in the habitable zone of M dwarfs may influence the construction of complex organic molecules.  However if such UV fluxes are detrimental to build complex organic molecules, below approximately nine meters of sea-water, this photo-dissociating and photo-ionizing radiation will not penetrate, allowing such complexity to flourish where there is a sufficient free energy gradient and raw materials, e.g., around hydtrothermal vents \citep[e.g.,][]{kiang2007}. 

\subsection{The effect of flare-driven atmospheric evolution on observation}
\label{sec:discuss_obs}

With missions on the horizon like the James Webb Space Telescope (JWST), and missions in planning such as LUVOIR, obtaining ever-increasing precision across the electromagnetic spectrum with transmission spectroscopy is certain - and of utmost importance to exoplanetary science and astrobiology. Given that we have a dataset of exactly one habitable planet, we need to develop understanding of how atmospheric evolution is driven by stellar activity to accurately interpret these spectra with respect to habitability and how atmospheres might be altered by the presence of life; of similar, related importance is correctly identifying abiotic false positive observations \citep{harman2015,schwieterman2016}. 

\begin{figure}
\centering
\includegraphics[width=0.8\textwidth]{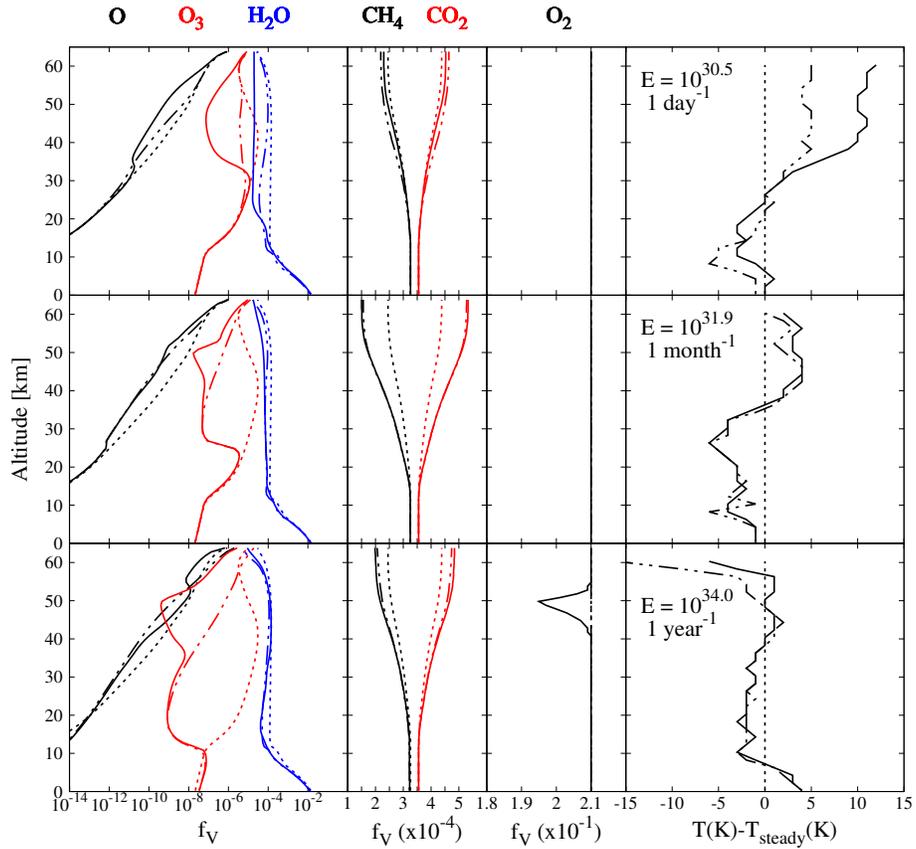}
\caption{Atmospheric profiles for the three parameter studies of 100 flares of log(E) 30.5 at frequency of 1 per day (top row), 31.9 with frequency of 1 per month (middle row), and 34.0 with frequency 1 per year(bottom row). In all cases, the dashed line represents initial steady state, the dash-dotted line signifies the state at the end of the flaring (before recovery), and the solid line represents the state at peak ozone loss during recovery.}
\label{fig:mixing_temp_100}
\end{figure}

\begin{figure}
\centering
\includegraphics[width=0.8\textwidth]{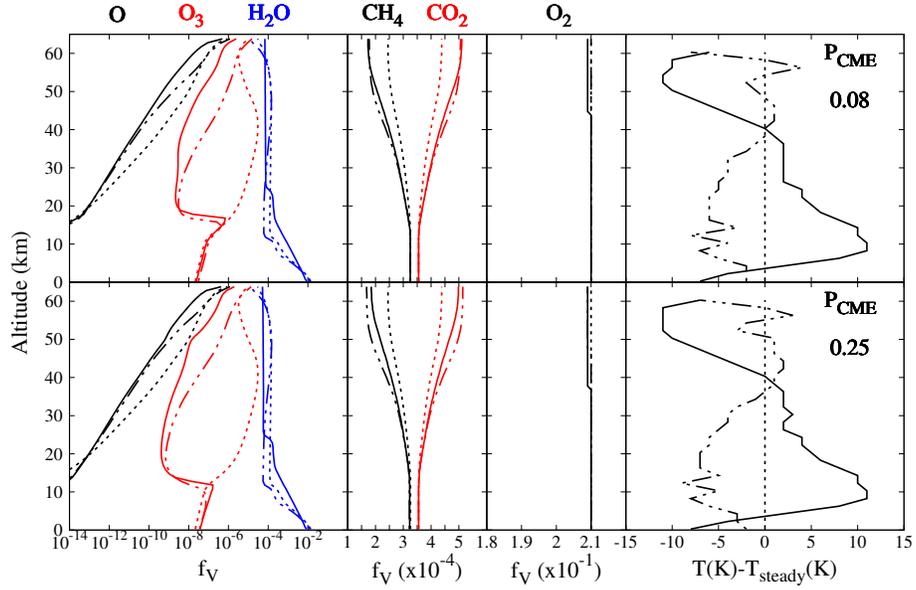}
\caption{Atmospheric profiles for two 10-year periods of flare activity, generated from the GJ1243 flare frequency distribution. The top row is with a conservative CME impact probability of 0.08, and the bottom row is the more permissive probability of 0.25. In all cases, the dashed line represents initial steady state, the dash-dotted line signifies the state at the end of the flaring (before recovery), and the solid line represents the state at peak ozone loss during recovery.}
\label{fig:mixing_temp}
\end{figure}

One particular strong atmospheric signal pointing to biotic chemical disequilibrium on Earth is the co-existence of abundant N$_2$, O$_2$ and H$_2$O \citep{krissansen2016}. In the present work, the levels of these particular species are not affected in either the 100 template flare simulations (from \S~\ref{sec:multi_emprot}, see Fig.~\ref{fig:mixing_temp_100}) or the 10 year real-CME simulations (from \S~\ref{sec:multi_emprot}, see Fig.~\ref{fig:mixing_temp}). However, it is noted that for our simulated starting point of an Earth-like planet, other species of interest that might be targeted by transmission spectroscopy can be altered significantly. For instance, the CO$_2$ in the upper atmosphere is notably increased by the photolysis of CH$_4$ driven by flaring activity, by $\geq$50 ppm at altitudes of 45 km and above, as seen in Figs.~\ref{fig:mixing_temp_100}~\&~\ref{fig:mixing_temp}. Similarly, the CH$_4$ levels have been depleted by similar amounts. O$_3$ obviously experiences significant reduction, and H$_2$O is altered.

As the more realistic simulations in Fig.~\ref{fig:mixing_temp} are artificially cut off after a maximum of 10 years, we do not show the result of the anticipated millions of years of stellar activity on the atmospheric state. In the lower panels of Fig.~\ref{fig:mixing_temp} one can see that the O$_2$ levels have been slightly eroded, by $\sim$0.1\%. While this is unlikely to produce significant alterations of spectral transit depth, continued reduction of O$_2$ over the periods of stellar activity exhibited by M dwarf hosts could potentially drive the signal below threshold for detection, giving a false indicator of the atmospheric state near the surface.

Our simulations indicate that a planet with an Earth-like atmosphere subjected to GJ1243 levels of flare and proton event activity is unlikely to be altered to the point of obscuring the strong N$_2$-O$_2$-H$_2$O chemical disequilibrium present. However, even on the short term of the present 10 year simulations, O$_3$ is rapidly eroded by multiple orders of magnitude, which indicates the possibility that no ozone layer may exist on such cases even if there is life producing oxygen, and there exists indication in the present simulations that O$_2$ could be further reduced by extended periods of M dwarf activity (see, e.g., Fig.~\ref{fig:mixing_temp_100}). Further work is required to explore the details of this particular consequence. 

\subsection{Other M dwarf hosts}

We have focused here on M dwarf activity observed on GJ1243. M dwarf stellar activity is highly variant, however, with GJ1243 only a moderately-active star. Other active hosts, such as GJ876, display a drastically lower level of activity and anticipated total proton flux \cite[e.g.][]{youngblood2017}. Even with the lower levels of event-specific proton flux estimated for GJ876 (10$^2$ -- 10$^3$ pfu compared to the events from 10$^4$ -- 10$^8$ in the present work), one finds that the long-term stability of O$_3$ is reduced on a relatively quick timescale, to $\sim$80-90\% of the equilibrium value after 40 months of flaring (see, e.g., Fig.~11 in \citet{youngblood2017}). Extrapolating that rate of loss leads to $\geq$99\% O$_3$ depletion after 10$^9$ -- 10$^{13}$ seconds ($\sim$10$^2$ -- 10$^5$ years), though as seen in Fig.~\ref{fig:davmultiprot}, it is possible the atmosphere reaches a new equilibrium O$_3$ column.

Similarly, there are M dwarfs capable of much higher levels of activity, with very active M3-M5 stars observed displaying levels of activity from 1-2 orders of magnitude higher than what we have used in the present work \citep[e.g.,][]{hilton2011}. Given the level of O$_3$ depletion seen in the present work, it is likely that planets orbiting more active hosts show a more rapid, more thorough destruction of any potential O$_3$ layer. 

\subsection{Recently observed Earth-like planets}

Recent observations have discovered potentially habitable worlds orbiting both Proxima Centauri and TRAPPIST-1, M6 and M8 stellar hosts, respectively \citep{anglada2016,gillon2017}. MOST observations have indicated that Proxima Centauri is highly active, with multiple flares per day at 10$^{29.5}$ erg, up to a 10$^{31.5}$ erg event every week \citep{davenport2016}. This level of activity could drive significant loss of O$_3$ (see, e.g., middle panel of Fig.~\ref{fig:davmultiprot}), especially given that the planet orbits at $\sim$0.0485 AU. The closer orbit increases the expected proton fluence and UV flux by an order of magnitude, compared to that in the present work --- which assumes the planet orbits at 0.16 AU. Balancing this is the lower energy flare distribution, which is lower by about an order of magnitude, which by the scaling used in the present work, roughly produces similar fluences at the planet as the orbital and energy scaling cancel each other - though the UV flux would still increase from the isotropic scaling. Therefore, the estimates in Fig.~\ref{fig:realmultiprot} can loosely be ascribed to an Earth-like atmosphere on Proxima Centauri b, but detailed modeling is required to obtain accurate results.

TRAPPIST-1 seems to currently exhibit infrequent flaring activity by optical flux measurements. \citet{vida2017} used K2 light curves to measure the flare activity of the star, and estimated events of order 10$^{30}$-10$^{33}$ erg occur from once every few days for the lower energy events, and approximately once per 100 days on the high end of the energy range. \citet{omalley2017} showed that the surface habitability of the planets in TRAPPIST-1's habitable zone are constrained by the activity level of the star and oxic state of the atmosphere. There are currently no constraints on the biologically relevant UV flux for TRAPPIST-1, but from Lyman $\alpha$ chromospheric observations and the optical observations above, it is likely the star is active in that wavelength range. 

If the surface habitability of these bodies, and those similar that are discovered in the future, is to be assessed, the effect of M dwarf flaring on any present atmosphere must be explored. In the present work, we assumed the starting point of an Earth-like atmosphere, which presupposes a high O$_2$ content and chemical disequilibrium driven by life. However, it is likely that observed planets consist of varied compositions that will require detailed modeling based on observation from future missions such as JWST or LUVOIR.

\subsection{Anoxic, Hazy Atmospheres}
\label{sec:haze}

It is important to recognize that hazy layers or the relatively high population of aerosols could change these findings, as they are efficient absorbers of UV radiation. In an early-Earth type anoxic atmosphere, \citet{arney2016} showed that the presence of haze can cut the incident UV-C radiation at the surface of the planet by a factor of $\sim$30 with a solar zenith angle of 60$^\circ$. While we cannot observe the conditions of the early Earth, we do have a decent laboratory in the solar system for the formation and evolution of atmospheric hazes: Saturn's satellite Titan. It has been observed that energetic ions from Saturn's magnetosphere can drive significant changes in ionospheric upper atmospheric chemistry \citep[e.g.][]{cravens2008}; such activity is thought to significantly contribute to the formation of high-amu organic haze layers that can then precipitate and coalesce into tholins at the surface of the satellite \citep[e.g.][]{waite2007,lavvas2013}.

Simulating the effects of stellar activity and proton chemistry on a terrestrial atmosphere after the development of photosynthesis but before the 'great oxidation event' could provide further insight into the shielding efficiency of aerosols and hazes during this early stage of the development of life, and therefore identify the amount of stress placed on an unshielded, early biosphere. Given increased activity during early solar evolution, and from M dwarfs in general, it is important to investigate how the presence of energetic proton precipitation would affect the formation and sustained chemistry of high altitude haze layers, and therefore their impact on UV-C radiation at the surface of a potentially habitable planet. 

\subsection{Unmagnetized Planet}
\label{sec:unmagnetized}

There is uncertainty in the ability for a global magnetic field to protect against the erosion of terrestrial planetary atmospheres by a stellar wind. On one hand, the magnetized Earth has retained a relatively thick atmosphere while (mostly) unmagnetized Mars has not; on the other hand, unmagnetized Venus has a $\sim$93 bar atmosphere and experiences approximately twice the solar wind energy deposition per unit area as Earth due to its closer orbital distance.

Similarly, there is a question related to the present work on the ability of a planetary magnetic field to protect the atmosphere from energetic proton deposition by the stellar wind. One might na\"{i}vely assume that the lack of a global field allows the protons from the stellar wind free reign to penetrate deeply into the upper atmosphere of such a planet. However, photoionization of the upper atmosphere leads to a spherical, conducting shell (ionosphere) that deflects the magnetized stellar wind. It has been suggested that for Venus that solar wind penetration isn't expected to drive significant chemical or physical changes to the ionosphere or lower layers of the atmosphere \citep[e.g.,][]{gombosi1980}. Interestingly, for planets without a global magnetic field that orbit close to their host stars, the increased XUV/EUV flux in the upper atmosphere are expected to produce an ionosphere of higher density and scale height, leading to deflection of the stellar wind at higher altitudes \citep[e.g.,][]{cohen2015}. However, this expansion could lead to more rapid atmospheric erosion, which over time could reduce the strength of the ionospheric deflection. The penetration of energetic protons, like those from CME/SEP events, are still an open question for potentially habitable exoplanetary atmospheres.

For an undisturbed Earth-like geomagnetic field, protons require an energy of $\geq$10 GeV to penetrate directly into the subtropical to equatorial ionosphere \citep[e.g.,][]{rodger2006}. While protons with these energies have been observed from high energy solar events, the indicate fluxes of $\geq$10 GeV protons are of factor $\geq$6 orders of magnitude lower than fluxes of protons of $\sim$100 MeV \citep[e.g.,][]{cliver2013}. However, even if these particles don't penetrate completely to the upper atmosphere, they can become temporarily trapped in the geomagnetic field, and energized protons flowing within the magnetosphere are capable of precipitation into the upper atmosphere, e.g., via pitch angle scattering. 

The penetration energy requirement also declines with increasing magnetic latitude for a dipole magnetic field \cite[e.g.,][]{shea1992,smart2000}. There are open regions at the separation between open and closed magnetospheric field lines, i.e. the cusp regions, where energetic protons could flow directly from the stellar wind into the upper atmosphere. Further, it is worth noting the geometry, orientation and magnitude of a planetary field; an inclined dipole field could allow direct proton access to the upper atmosphere, and a strong quadrupole moment, for instance, has polar regions similar to the dipole field geometry of Earth, but also a circum-equatorial region with intersecting field lines where precipitation could occur. 

The interactions between stellar wind, and energetic stellar outflows, are complex combinations of magnetic star-planet interactions that require further study. This begs the question of how effective magnetic fields are at deflecting (non-)relativistic proton events, shielding the planetary atmosphere. We plan to investigate and address these issues in Part 2, in an companion paper to the present work. 

\section{Conclusions}

The conditions under which life can develop are not certain, but the clement conditions must be sufficiently stable. For M dwarf orbiting planets, it is clear they will experience a high flux of electromagnetic and particle radiation - understanding these effects on the atmosphere are imperative to predict the likelihood of surface life. As shown in the present work, even with conservative assumptions, M dwarf activity can drive extreme loss in the O$_3$ column for an Earth-like planet within a period of time that is several orders of magnitude shorter than the known active lifetimes for low mass stars. Absent other shielding, short wavelength UV-B and UV-C will blanket the surface. The effects of such radiation can be harmful to complex organic structures, but there is much study to be done to reproduce such an environment to understand how life could emerge.

In particular, we find:

\begin{itemize}
    \item EM-only flaring for a stellar host like GJ1243 is unlikely to drive significant O$_3$ loss, with only $\sim$37\% O$_3$ column loss by the present age of the universe for the modeled flaring activity. In contrast, the same loss can be achieved in a minimum $\sim$1.3 yr by weekly proton impacts from the lowest energy events simulated.
    \item O$_3$ column loss by flaring parameters from an observed host like GJ1243, as shown in Fig.~\ref{fig:realmultiprot}, can drive $\geq$99.9\% destruction of the O$_3$ column in as little as 0.5 Myr, with a trend of increasing loss that is unlikely to reverse.
    \item The impact of one ADLeo sized flare per month is sufficient to drive a loss of 99.99\% of O$_3$ column within a $\sim$8 yr. In this case, the surface of a planet will experience {\it quiescent} UV-C flux of $\sim$0.18 W m$^{-2}$, with subsequent 10$^{34}$ flares delivering $\sim$127 kJ m$^{-2}$ of UV-C for a single event. Further experiments are needed to evaluate the impact of these fluxes on the onset of complex organic chemistry and life. 
    \item In addition to reducing the O$_3$ mixing ratio by orders of magnitude, our results suggest that atmospheric species of interest (e.g., CO$_2$, CH$_4$) can be modified to levels that will be observable by upcoming missions. Mixing ratio vertical profiles change slightly for steady state conditions, while the high level of stellar activity can drive long-term profile changes of at least ~$\sim$100 ppm within the span of decades.
\end{itemize}

Future efforts should include detailed simulations of varying magnetospheric conditions on deflection of proton events to gain an understanding of how well M dwarf orbiting planets are protected by an intrinsic magnetic dipole. Of further interest is developing simulations of atmospheric effects of M dwarf activity for planets with non-Earth-like starting conditions, e.g., high CO$_2$, high O$_2$. 

\begin{acknowledgements}
M.T. would like to thank Erika Harnett, Joshua Krissansen-Totten and Eddie Schwieterman for helpful discussion, and acknowledges the generous support of the NAI Virtual Planetary Laboratory and University of Washington Astrobiology Program. This material is based upon work performed by the NASA Astrobiology Institute's Virtual Planetary Laboratory Lead Team, supported by the National Aeronautics and Space Administration through the NASA Astrobiology Institute under Cooperative Agreement number NNA13AA93A. We acknowledge use of NASA/GSFC's Space Physics Data Facility's OMNIWeb (or CDAWeb or ftp) service, and OMNI data. We also acknowledge the use of GOES data at NOAA's National Centers for Environmental Information (NCEI) Space Weather data access.

A.S. also acknowledges support from UNAM PAPIIT project IN109015.
\end{acknowledgements}

\clearpage
\section{Figures}

\clearpage

\clearpage

\clearpage

\clearpage

\clearpage

\clearpage
 
\clearpage
 
\clearpage

\clearpage

\clearpage

\clearpage

\clearpage

\clearpage

\clearpage


\begin{deluxetable}{lccc}
\tablewidth{0.75\textwidth}
\tablecaption{Henry coefficients, mixing @ ground}
\tablenum{1}
\tablehead{\colhead{Species} & \colhead{k$_H$} & \colhead{d(ln(k$_H$))/d(1/T)} & \colhead{f (z=0) } \\
\colhead{} & \colhead{[ mol kg$^{-1}$ bar$^{-1}$]} & \colhead{[K]} & \colhead{[arb]} } 
\startdata
O$_{2}$ & 1.3$\times$10$^{-3}$ & 1500 & 0.21 \\
CO$_{2}$ & 3.55$\times$10$^{-2}$ & 2400 & 3.55$\times$10$^{-4}$ \\ 
N$_{2}$ & 6.25$\times$10$^{-4}$ & 1300 & 0.78
\enddata
\tablecomments{Henry coefficients and surface mixing ratios for each unfixed species. \label{tab:henry_coeff}}

\end{deluxetable}

 \clearpage

\begin{deluxetable}{lcc}
\tablewidth{0.6\textwidth}
\tablecaption{Time estimation to $\geq$~99\% O$_3$ loss, by interflare separation}
\tablenum{2}
\tablehead{\colhead{} & \colhead{10$^{31.9}$ erg} & \colhead{10$^{34.0}$ erg}}
\startdata
2 hour sep &  3.50$\times$10$^7$  & 1.89$\times$10$^7$ \\
1 day sep &  7.38$\times$10$^7$ & 3.22$\times$10$^7$ \\
1 week sep &  2.55$\times$10$^8$ & 8.33$\times$10$^7$ \\
1 month sep &  {\bf 2.64$\times$10$^{14}$} & 1.35$\times$10$^{8\,**}$\\
Total fluence [prot cm$^{-2}$] & 1.82$\times$10$^{12}$ & 8.49$\times$10$^{14}$ \\
\enddata
\tablecomments{Time to $\geq$~99\% O$_3$ loss assuming continued flare activity beyond the 100 flares simulated as shown in Fig.~\ref{fig:davmultiprot}. Power law fits to O$_3$ loss rate for last $\sim$20 flares in each case, and are shown for the most likely cases for a planet orbiting GJ1243. Times in bold represent the most likely time to loss for a planet orbiting GJ 1243 in the HZ for conservative CME/SEP geometries (see \S~\ref{sec:proton_events}). \\ ** - For the 10$^{34}$ erg flares with 1 month interflare separation, the system sustained O$_3$ losses $\geq$~99\% during the flaring period at this time above, no fit necessary.\label{tab:o3_fits}}

\end{deluxetable}

\clearpage

\begin{deluxetable}{lccc}
\tablewidth{\textwidth}
\tablecaption{ Integrated UVC flux (see Fig.~\ref{fig:uvflux})}
\tablenum{3}
\tablehead{\colhead{} & \colhead{UVC @ TOA} & \colhead{UVC @ Surface} & \colhead{Integrated UVC} \\
\colhead{} & \colhead{W m$^{-2}$} & \colhead{W m$^{-2}$} & \colhead{J m$^{-2}$} }
\startdata
Earth & 6.73 & 2.13$\times$10$^{-14}$ & -\\
\emph{GJ1243 Planet} \\
Quiescence, full O$_3$ & 2.76 & 4.15$\times$10$^{-13}$ & - \\
Quiescence, 0.1 O$_3$ & 2.76 & 1.05$\times$10$^{-6}$ & - \\
Quiescence, 0.08 O$_3$ & 2.76 & 1.45$\times$10$^{-5}$ & - \\
Quiescence, 1.6$\times$10$^{-4}$ O$_3$ & 2.76 & 0.18 & - \\
10$^{31.9}$ peak, 0.08 O$_3$ & 11.4 & 1.7$\times$10$^{-5}$ & 4.5$\times$10$^{-3}$ \\
10$^{33.6}$, 0.08 O$_3$ & 180.4 & 2.98$\times$10$^{-4}$ & 4.31$\times$10$^{-1}$ \\
10$^{30.5}$ peak, 1.6$\times$10$^{-4}$ O$_3$ & 3.63 & 0.24 & 3.53$\times$10$^{1}$ \\
10$^{34}$ peak, 1.6$\times$10$^{-4}$ O$_3$ & 368.76 & 60.8 & 1.27$\times$10$^5$
\enddata
\tablecomments{UVC flux ($<$2800 \AA) in W m$^{-2}$ - integrated over the examples in Fig.~\ref{fig:uvflux} - at the top of atmosphere (TOA) and the planetary surface for quiescent conditions with steady state O$_3$ column, quiescent conditions with depleted O$_3$ column (0.1, 0.08, and 1.6$\times$10$^{-4}$ of equilibrium value), and peak values for 10$^{30.5}$, 10$^{31.7}$, 10$^{31.9}$, and 10$^{34}$ erg flares with associated depleted O$_3$ values. \label{tab:uvflux}}

\end{deluxetable}

 \clearpage


\bibliography{bib}

\end{document}